\documentclass[aps,superscriptaddress,prc,twocolumn,nofootinbib]{revtex4}

\usepackage{amsmath,bm}
\usepackage{graphicx}
\usepackage{epstopdf}
\usepackage{subfigure}
\usepackage{epsfig}
\usepackage{amsmath,amssymb,amsfonts}
\usepackage{color}
\usepackage[utf8]{inputenc}
\usepackage{verbatim}
\usepackage{array}
\allowdisplaybreaks
\def\bea#1\eea{\begin{align}#1\end{align}}
\def \be  {\begin{equation}}
\def \ee  {\end{equation}}

\newcommand{\bef}{\begin{figure}[t]\centering}
\newcommand{\eef}{\end{figure}}

\begin{document}

\renewcommand{\topfraction}{0.85}
\renewcommand{\textfraction}{0.1}
\renewcommand{\floatpagefraction}{0.75}

\date{\today  \hspace{1ex}}

\title{Quenching of jets tagged with $W$ bosons in high-energy nuclear collisions}

\author{Shan-Liang Zhang}
\affiliation{Key Laboratory of Quark \& Lepton Physics (MOE) and Institute of Particle Physics,
 Central China Normal University, Wuhan 430079, China}
\affiliation{Guangdong Provincial Key Laboratory of Nuclear Science, Institute of Quantum Matter, South China Normal University, Guangzhou 510006, China.}
\affiliation{Guangdong-Hong Kong Joint Laboratory of Quantum Matter, Southern Nuclear Science Computing Center, South China Normal University, Guangzhou 510006, China.}

\author{Xin-Nian Wang}
\affiliation{Nuclear Science Division, Lawrence Berkeley National Laboratory, Berkeley, CA 94720, USA \footnote{current address}}
\affiliation{Key Laboratory of Quark \& Lepton Physics (MOE) and Institute of Particle Physics,
 Central China Normal University, Wuhan 430079, China}

\author{Ben-Wei Zhang \footnote{bwzhang@mail.ccnu.edu.cn}}
\affiliation{Key Laboratory of Quark \& Lepton Physics (MOE) and Institute of Particle Physics,
 Central China Normal University, Wuhan 430079, China}
\affiliation{Guangdong Provincial Key Laboratory of Nuclear Science, Institute of Quantum Matter, South China Normal University, Guangzhou 510006, China.}
\affiliation{Guangdong-Hong Kong Joint Laboratory of Quantum Matter, Southern Nuclear Science Computing Center, South China Normal University, Guangzhou 510006, China.}

\begin{abstract}
    We carry out the first detailed calculations of jet production associated with $W$ gauge bosons in Pb+Pb collisions at the Large Hadron Collider (LHC). In our calculations, the production of $W$+jet in p+p collisions as a reference is obtained by Sherpa, which  performs next-to-leading-order matrix element calculations matched to the resummation of parton shower simulations, while jet propagation and medium response in the quark-gluon plasma are simulated with the Linear Boltzmann Transport (LBT) model. We provide numerical predictions on seven observables of $W$+jet production with jet quenching in Pb+Pb collisions: the medium modification factor for the tagged jet cross sections $I_{AA}$, the distribution in invariant mass between the two leading jets in $N_{jets}\ge 2$ events $m_{jj}$, the missing $p_T$ or the vector sum of the lepton and jet transverse momentum  $|\vec{p}_T^{Miss}|$, the summed scalar $p_T$ of all the jets in an event  $S_T$,   transverse momentum imbalance  $x_{jW}$, average number of jets per $W$ boson $R_{jW}$, and azimuthal angle between the $W$ boson and jets $\Delta \phi_{jW}$. The distinct nuclear modifications of these seven observables in Pb+Pb relative to that in p+p collisions are presented with detailed discussions.

\end{abstract}


\maketitle

\section{introduction}
\label{sec:Intro}
Jet quenching due to strong interaction between  energetic partons and the dense QCD medium has long been proposed as an excellent hard probe of the properties of the quark-gluon-plasma (QGP) created in relativistic heavy-ion collisions (HICs)~\cite{Wang:1991xy,Gyulassy:2003mc,Qin:2015srf,Vitev:2008rz,Vitev:2009rd,Qin:2010mn,CasalderreySolana:2010eh,Young:2011qx,He:2011pd,ColemanSmith:2012vr,Zapp:2012ak,Ma:2013pha,Senzel:2013dta,Casalderrey-Solana:2014bpa,Milhano:2015mng,Chang:2016gjp,Majumder:2014gda,Chen:2016cof, Chen:2016vem,He:2020iow,Chien:2016led,Apolinario:2017qay,Connors:2017ptx,Dai:2018mhw,Wang:2019xey,Chen:2019gqo,Yan:2020zrz,Chen:2020pfa}. Among a wealth of jet quenching observables, the production of jets in association with a gauge boson has been regarded as a `golden channel' due to  some of  its unique features~\cite{Wang:1996yh,Wang:1996pe}. Since gauge bosons produced in the initial hard scattering do not participate in strong interactions with the medium, the transverse momentum of the boson closely reflects the initial energy of the  leading jets before they interact with the medium. In addition, jets associated with a gauge boson are dominated by quark jets, which can help constrain  the flavor dependence of parton energy loss.
Recently $\gamma/$Z -jet correlations~\cite{Qin:2009bk,Dai:2012am,Wang:2013cia,Casalderrey-Solana:2015vaa,KunnawalkamElayavalli:2016ttl,Kang:2017xnc,Chen:2018fqu,Luo:2018pto,Neufeld:2010fj,Neufeld:2012df,Zhang:2018urd,Chen:2017zte,Chen:2020tbl,Zhang:2018kjl,Sirunyan:2017qhf,Sirunyan:2017jic,Chatrchyan:2013tna}, and H+jet processes~\cite{Berger:2018mtg,Chen:2020kex} have already been investigated within several theoretical models and by experiments in Pb+Pb collisions at $\sqrt s=5.02$ TeV.  In this paper, we perform a first quantitative study of $W$+jets in high-energy heavy-ion collisions.

At the leading order (LO) in perturbation theory,  jet production associated with $W$ bosons attributes  mainly to two subprocesses: quark-antiquark annihilation $q\bar q\prime\rightarrow W g$ and Compton process $qg\rightarrow W q$. At very large momentum transfer, the Compton process dominates, therefore, jets recoiling from a $W$ boson are predominately quark jets. In this respect, $W$+jet can help further constrain  the flavor dependence of jet quenching.  When a $W$ boson is produced in the center of mass frame at LO, its momentum component transverse to the beam axis is balanced by a back-to-back jet with the same momentum in the transverse plane, resulting in the divergence of transverse momentum imbalance at $x_{jW}=p_T^{jet}/p_T^{W}\simeq1$ and azimuthal angle correlation at $\Delta\phi_{jW}=|\phi_{jet}-\phi_{W}|=\pi$ \cite{Luo:2018pto,Chatrchyan:2013tna,Zhang:2018urd}.  With higher-order perturbative corrections\cite{Boughezal:2016dtm,Czakon:2020coa,Boughezal:2015dva,Kallweit:2015dum}, additional hard and soft gluon radiations may affect the $W$-jet correlations, for instance, the balance of transverse momentum is smeared and azimuthal angle correlation is broadened.

 As in $Z$+jets \cite{Zhang:2018urd,Zhang:2018kjl,Chatrchyan:2013tna}, the next-to-leading-order (NLO) calculation  does not take the resummation of soft/collinear radiation into account and has only limited number of finial particles. Though the NLO calculation can describe  transverse momentum spectra of jets, it is,   however,  insufficient to study $W$-jet correlations in azimuthal angle which suffer from divergence in  the large angle region.
Monte Carlo (MC) event generator PYTHIA which employs leading-order matrix element (ME) merged  with parton shower (PS) contains some high-order corrections from both real and virtual contributions. It is, however, short of additional hard or wide-angle radiations from high-order matrix element calculations.
Simulations matching the NLO  with PS \cite{Zhang:2018urd,Zhang:2018kjl,Chatrchyan:2013tna,Sun:2018icb,Chien:2019gyf,Kang:2019ahe}, on the other hand,  provide a satisfactory description of a wide variety of experimental observables of $W^{\pm}/Z/\gamma$+ jet in the whole phase space.  Therefore, we will utilize an improved reference of gauge boson tagged jet production in proton-proton (p+p) collisions to study $W$-jet correlations in relativistic heavy-ion collisions (HIC) at the Large Hadron Collider (LHC).

In this paper, we will carry out a first systematic study of $W$+jets in both  p+p and heavy-ion collisions (Pb+Pb) at $\sqrt{s_{\rm NN}}=5.02 $ TeV.  With MC event generator SHERPA~\cite{Gleisberg:2008ta}, which can perform NLO matrix element calculations matched to the resummation of parton showers, we provide excellent  baselines of p+p collisions at 5.02 TeV.  We will then use the Linear Boltzmann Transport (LBT) model ~\cite{Li:2010ts,He:2015pra, Cao:2016gvr} to simulate jet propagation and medium response and predict the medium modifications of several specific observables of $W$+jet in HIC:  the distribution in invariant mass between the two leading jets in $N_{jets}\ge 2$ events $m_{jj}$, the medium modification of jet spectra in different $p_T^W$ intervals, the modification of the distributions in $|\vec{p}_T^{Miss}|$ which is the vector sum of the lepton and jets transverse momentum, and the summed scalar $p_T^{jets}$ of all the jets in the event $S_T$. We will also provide numerical results for several familiar observables of $W$-jet correlations in HIC, which have been utilized to investigate $Z$+jet in HIC~\cite{Zhang:2018urd}: the shift of the transverse momentum imbalance of $W$+jet as well as its mean value between p+p and Pb+Pb collisions, the modification of azimuthal angle correlations, and the number of tagged jets per $W$ boson.

The rest of the paper is organized as follows. In Sec.~\ref{sec:framework} we present the framework for the calculation of jet production in association with $W$ bosons both in p+p and Pb+Pb collisions. We also describe how jets tagged by a $W$ boson are produced  in SHERPA and transported in LBT.
In Sec.~\ref{sec:results} we present medium modifications of seven observables of $W$+jet in Pb+Pb relative to that in p+p collisions.
In Sec.~\ref{sec:conclusion} we summarize our study.

\section{Framework description}
\label{sec:framework}

\subsection{$W$+jets in p+p collisions at NLO with PS }

In our calculations,  jet productions in association with a $W$ boson in p+p collisions are simulated within a MC event generator SHERPA 2.24~\cite{Gleisberg:2008ta}, which can perform NLO ME calculations  matched to the PS with several merging schemes. AMEGIC++~\cite{Krauss:2001iv} and COMIX~\cite{Gleisberg:2008fv} are SHERPA's original matrix-element generators which provide tree-level matrix-elements and create the phase-space integration as well. MC programs OpenLoops~\cite{Cascioli:2011va} is interfaced with SHERPA to provide the virtual matrix-elements.  MEPS@NLO merging method~\cite{Hoeche:2009rj,Hoche:2010kg,Hoeche:2012yf} is used to yield improved matrix elements for multiple jets production at NLO  matched to the resummed parton showers~\cite{Gleisberg:2007md,Schumann:2007mg}. LHAPDF is interfaced with SHERPA and the parton distribution  function (PDF)set `CT14 NLO'~\cite{Hou:2016sho} is used to provide the PDF for partons that participate in the hard interaction in p+p collisions.

\begin{figure}
\includegraphics[width=0.56\textwidth]{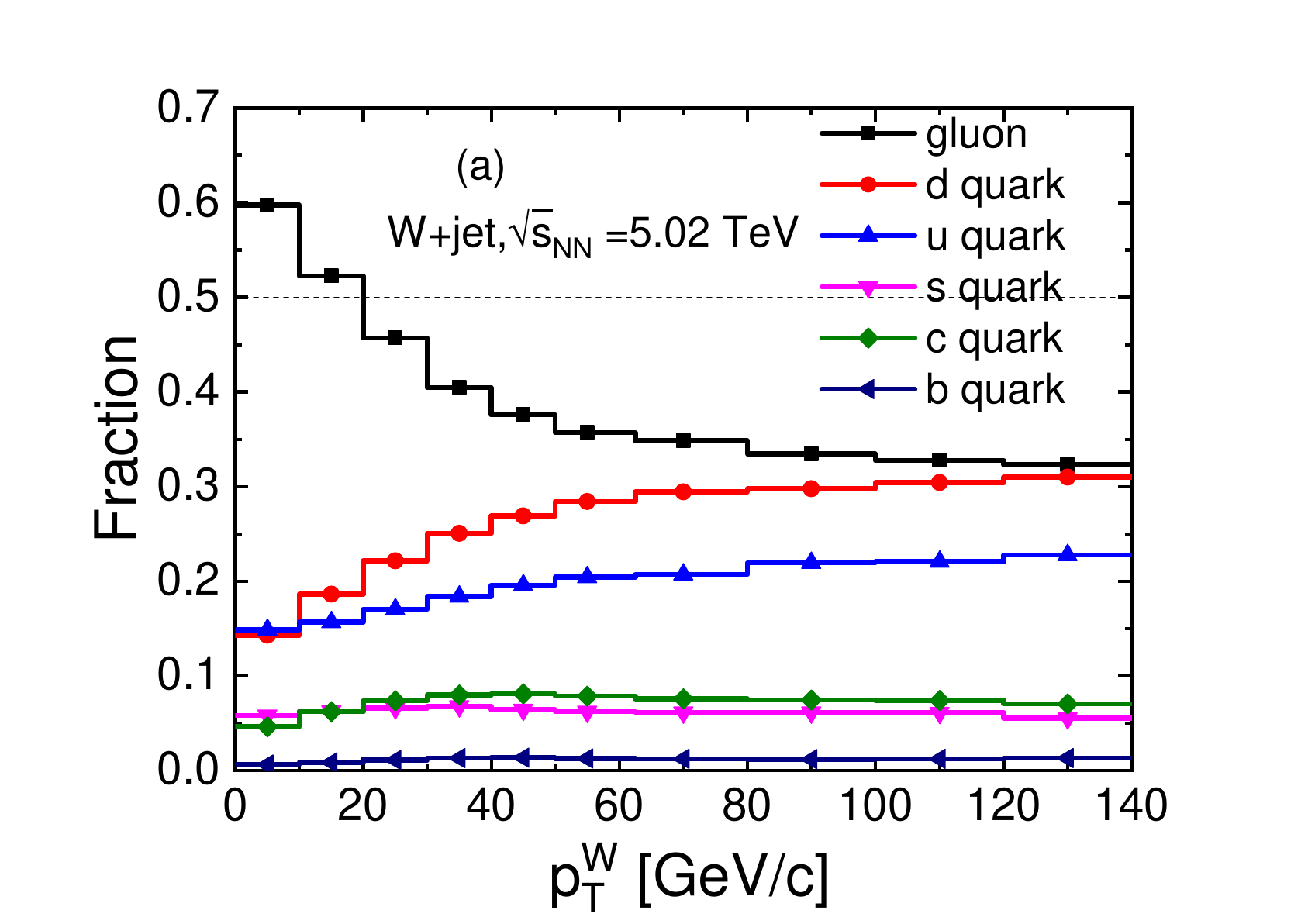}
\includegraphics[width=0.56\textwidth]{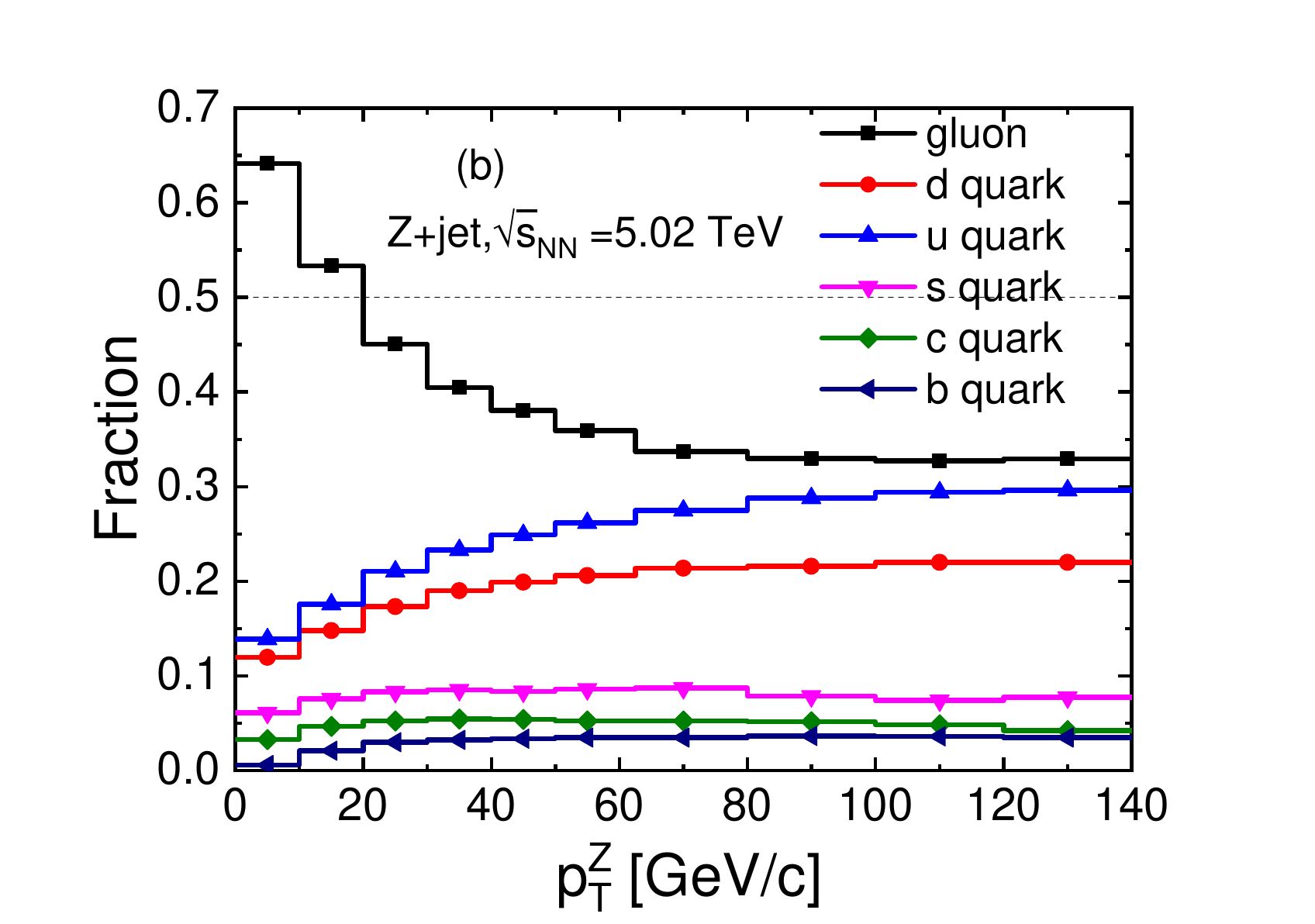}

  \caption{(Color online) Fraction of leading parton flavor triggered by: (a) $W$ boson  and (b) $Z$ boson as a function of  gauge boson transverse momentum $p_T^V$.   }\label{flavor}
\end{figure}

Though $W$ and $Z$ bosons have a lot in common, $W$ bosons have some unique properties as compared to $Z$ bosons. For instance, $W$ bosons carry  electric charge and would change the flavor of the jet parton. As a result, the flavors of jets associated with W's are different from a $Z$ boson.  We have classified the flavor of the leading parton triggered by a $Z$ boson and $W$ bosons in Fig.~\ref{flavor}. As can be seen, the quark fractions  increase significantly with the transverse momentum of the gauge boson. At high gauge boson energy, the jets are dominated by quark jets.  Compared to jet production associated  with a $Z$ boson, the fraction of the leading parton flavor tagged with $W$ bosons is quite different. For instance, the fraction of $u$ quark associated with $W$ bosons is almost the same as the fraction of $d$ quark associated with a $Z$ boson and the fraction of $s$ quark associated with a $Z$ boson is almost the same as the fraction of $c$ quark associated with a $W$ boson as a result of isospin symmetry. The different parton  flavor fractions would lead to different fractions of hadrons as well as different jet properties. The comparison between $W$+jet and $Z$+jet would provide new opportunities to explore the jet tomography of the QGP.  The difference  in the flavor fraction is a result of different production mechanisms of $W$ and $Z$ boson in hard scattering. In this regard, W's associated with jets or hadron can be used to constrain non-perturbative hadronization models in both p+p and Pb+Pb collisions as well as flavor dependence of jet quenching, which is beyond the scope  of this paper and will be discussed in the future.

  In order to compare with experimental measurements, we select the $W$ bosons and associated jets according to the kinematic cuts adopted by ATLAS experiment~\cite{Aad:2014qxa}. The electrons are constrained in the phase space $p_T > 25$ GeV/$c$ and $|\eta| < $  2.47 and are rejected in the transition region ($1.37 < |\eta| < 1.52$).  Muons are required to have $p_T >25 $ GeV/$c$ and $|\eta|<2.4$. Additionally, jets are reconstructed using the anti-kt algorithm~\cite{Cacciari:2011ma,Cacciari:2008gp} with a radius parameter $R=\sqrt{\Delta \eta^2+\Delta\phi^2}$ = 0.4 using all the final  state partons. Jets are required to have  $p_T > 30$ GeV/$c$ and $|y| < 4.4$ and are removed if a jet is within $\Delta R$ = 0.5 of an electron or muon. Furthermore, since the $W$ boson eventually decays into an electron and a neutrino, events are required to have significant missing transverse momentum and large transverse mass to
compensate the  missing information of the neutrino which can not be detected directly by experiment.
The missing transverse momentum is defined as the negative vector sum of the transverse momentum of leptons, photons, and jets as well as the soft deposits in the calorimeter, and is required to have $E^{miss}_T =-|\vec{p}_T^{\ l}+\vec{p}_T^{\gamma}+ \sum \vec{p}_T^{jets}+\vec{p}_T^{\ soft}|> 25$ GeV/$c$~\cite{Aad:2014qxa,Aad:2019sfe}.
Transverse mass is defined as $m_T=\sqrt{ 2p_T^lp_T^\nu(1-\cos(\phi^l-\phi^\nu))}$ and required to have $m_T > 40$ GeV/$c$.

The differential cross section of jet production  associated with a $W$ boson as a function of jet transverse momentum calculated by SHERPA is compared with the experimental data~\cite{Aad:2014qxa} in Fig.~\ref{ppbaseline}a. The distribution in di-jet invariant mass $m_{jj}=\sqrt{(E^L+E^{SubL})^2-(\vec{p}^{L}+\vec{p}^{SubL})^2}$ between the two leading jets in $N_{jets}\ge $ 2 events is also calculated and compared with experimental data in  Fig.~\ref{ppbaseline}b.  The  jet distributions in association with a $W$ boson production from SHERPA show excellent agreement with the experimental data and can be used as references and inputs for the energy loss models to study jet-medium interactions  in heavy-ion collisions. The jet spectrum monotonically decreases as a function of jet transverse momentum.  However, the distribution as a function of the dijet invariant mass increases when $m_{jj}< M_W$ and decreases steeply when $m_{jj}>M_W$, which is similar with the $W$ boson mass distribution and  quite different from inclusive dijet mass distribution which is a monotonic function of $m_{jj}$.

\bef

\includegraphics[width=0.56\textwidth]{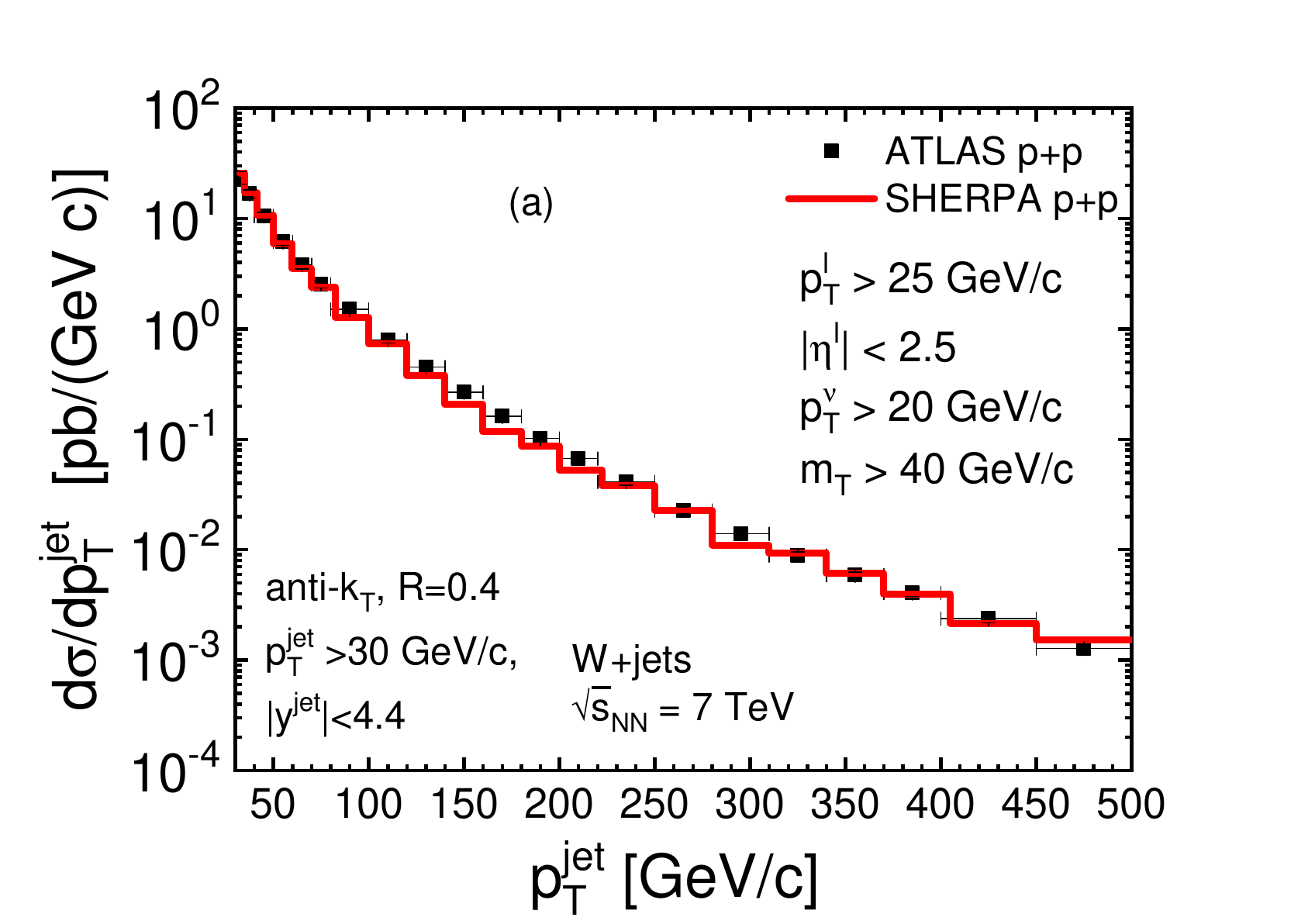}
\includegraphics[width=0.56\textwidth]{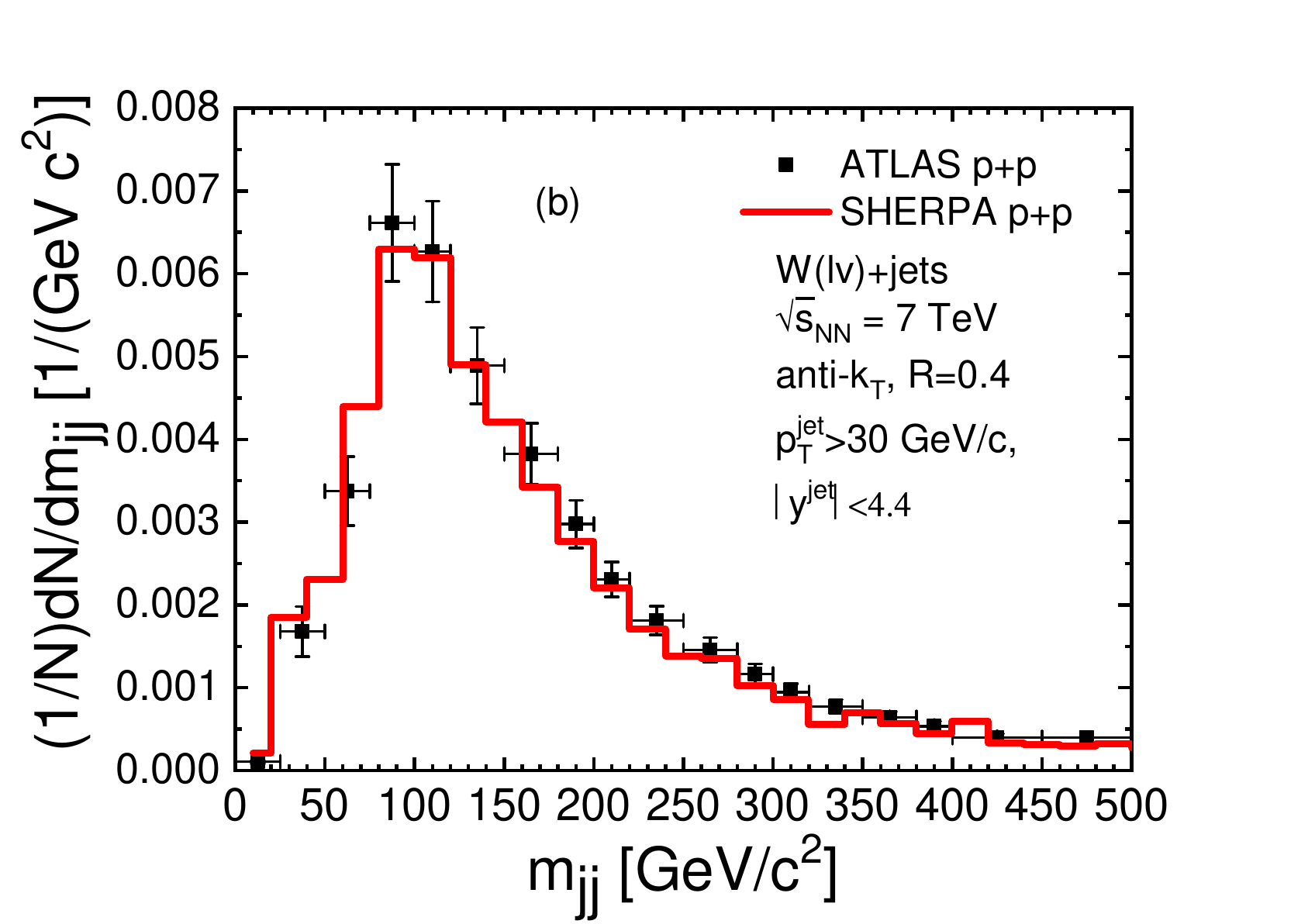}

\vspace{-0pt}
\caption{(Color online) (a) Differential cross section for the production of  $W$+jets as a function of the transverse momentum of the associated jets at $\sqrt {s_{NN}} $= 7 TeV and the comparison with the ATLAS experimental data (black).  (b) Normalized distributions of events passing the $W$ + jets selection cut as a
function of the dijet invariant mass $m_{jj}$ between the two leading jets in $N_{jets}\ge $ 2 events and the comparison with the ATLAS experimental data (black).}
\label{ppbaseline}
\eef

\subsection{$W$+jet in Pb+Pb collisions within the LBT model}

\begin{figure}
  \centering
  \includegraphics[scale=0.35]{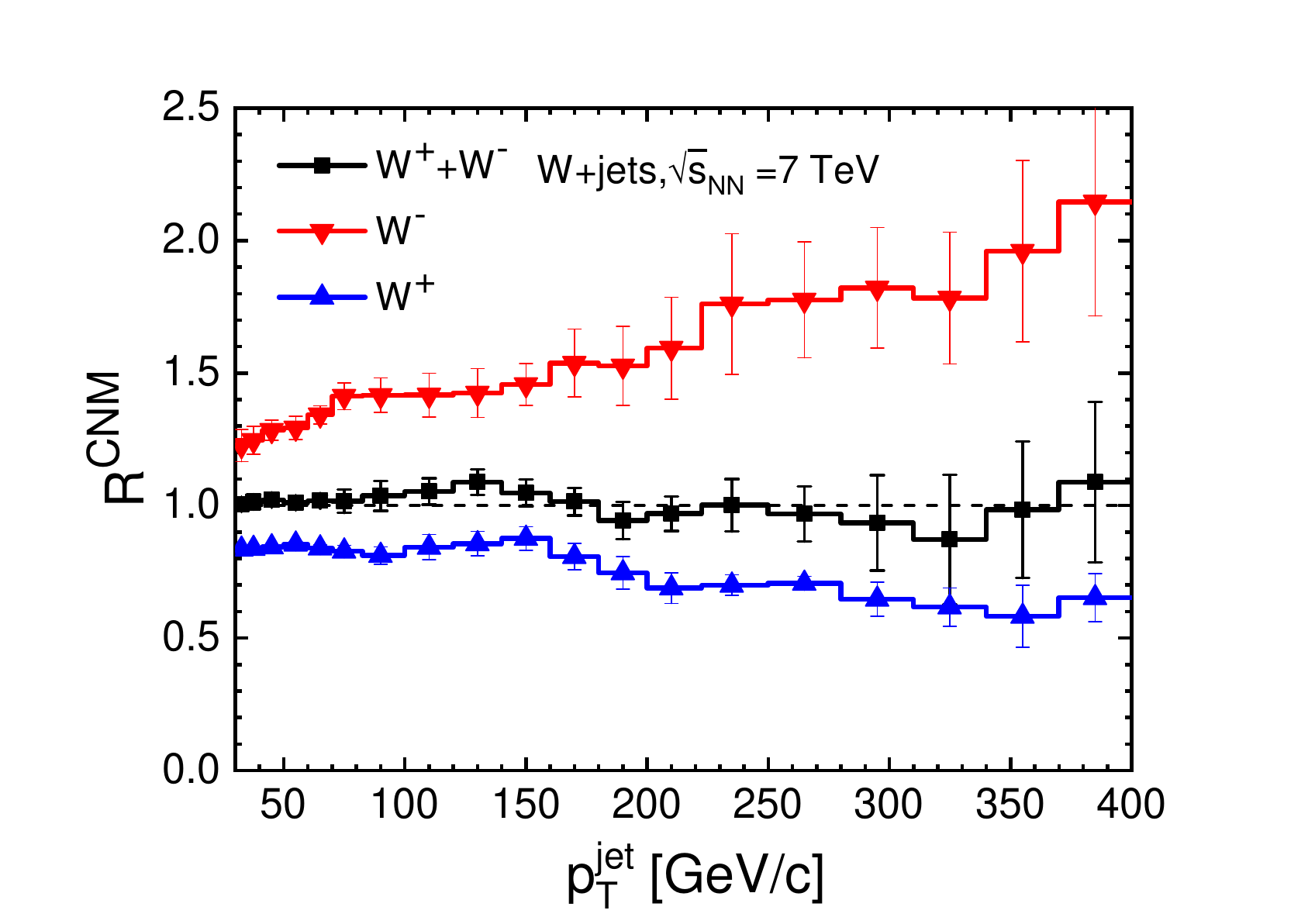}
   \vspace{5pt}
  \caption{(Color online)  The modification of jet spectrums tagged by  $W^+$,$W^-$ and $W^++W^-$  due to the cold nuclear effect with EPPS16 modified CT14nlo pdf set.}\label{wjetpp}
\end{figure}

For a quantitative investigation of the jet properties associated with a vector gauge boson in heavy-ion collisions, cold nuclear matter (CNM) effects should also be taken into consideration. In our calculations, we use the EPPS16~\cite{Eskola:2016oht} nuclear parton distribution functions (nPDF's)  in  the LHAPDF library to investigate the cold nuclear matter effect due to nuclear modification of the parton distribution functions. Cold nuclear matter effects are negligible  in the distribution of gamma/$Z$+jets in the kinematic ranges we are interested in. However, the cross section of jet production associated with a $W$ boson is rather sensitive to the isospin dependence of nPDF due to the production of the charged $W$ gauge bosons.  The nuclear modification factors for the jet $p_T$ distribution due to CNM  are calculated and shown in  Fig.~\ref{wjetpp}. As one can see, $W^-$ is enhanced by 20$\%$ while $W^+$ is suppressed  by 20$\%$  due to the isospin dependence of nPDF in Pb nuclei.  However, the CNM effect beyond the isospin dependence of nPDF is negligible as seen in the modification factor for the sum of $W^+$ and $W^-$ triggered jets.  Similar conclusions are reached in \cite{Ru:2014yma}. Therefore, the isospin dependence of nPDF must be taken into account for the study of nuclear modification of jet production associated with $W^+$ or  $W^-$  bosons. However, CNM effect becomes negligible when the final results are averaged over $W^+$ and  $W^-$.

 In our study,  jet propagation, parton energy loss and the medium response in hot/dense QCD medium due to jet-medium interactions are simulated within the Linear Boltzmann transport (LBT) model~\cite{Li:2010ts,He:2015pra, Cao:2016gvr}, which is based on the  Boltzmann equation,
\begin{equation}\begin{split}
 p_1\cdot\partial f_a(p_1)&=-\int\frac{d^3p_2}{(2\pi)^32E_2}\int\frac{d^3p_3}{(2\pi)^32E_3}\int\frac{d^3p_4}{(2\pi)^32E_4}\\
&\sum _{b(c,d)}[f_a(p_1)f_b(p_2)-f_c(p_3)f_d(p_4)]|M_{ab\rightarrow cd}|^2\\
&\times S_2(s,t,u)(2\pi)^4\delta^4(p_1+p_2-p_3-p_4),
 \end{split}\end{equation}
for parton propagation in the QGP medium, where $f_i$'s are  phase-space distributions of medium parton.
 $S_2(s,t,u)$ is Lorentz-invariant regulation condition to regulate all soft and collinear divergency.
 Elastic scatterings are simulated with the corresponding matrix elements $|M_{ab\rightarrow cd}|$ which include the complete set of leading order  $2\rightarrow 2$ elastic scattering processes.

The induced gluon radiation from inelastic scattering is numerically incorporated into LBT according to the High-Twist formalism~\cite{Guo:2000nz,Zhang:2003yn,Zhang:2003wk}:
\begin{equation}
\frac{dN_g}{dxdk_\perp^2 dt}=\frac{2\alpha_sC_AP(x)\hat{q}}{\pi dk_\perp^4}\left(\frac{k_\perp^2}{k_\perp^2+x^2M^2}\right)^2\sin^2\left(\frac{t-t_i}{2\tau_f}\right),
 \end{equation}
 where, $x$ and $k_\perp$ are the energy fraction and transverse momentum of the radiated gluon, respectively,
 $P(x)$ is the splitting function, $\hat{q}$ is the jet transport coefficient which is calculated from the elastic scattering, and
 $\tau_f=2Ex(1-x)/(k_\perp^2+x^2M^2) $ is the formation time of the radiated gluon. The medium information is provided by 3+1D CLVisc hydrodynamics~\cite{Pang:2012he,Pang:2014ipa} with the initial condition provided by the AMPT~\cite{Lin:2004en} Monte Carlo model.
 LBT has been successful in describing experimental data  on the suppression of large $p_T$ hadrons~\cite{He:2015pra, Cao:2016gvr}, inclusive jets~\cite{He:2018xjv}, $\gamma$-hadron/jets~\cite{Chen:2017zte,Chen:2020tbl,Luo:2018pto} correlations and $Z$+jet production~\cite{Zhang:2018urd}.

\section{Numerical results and discussions}
\label{sec:results}

In this section, we will present predictions about the modifications of $W$+jets event distributions and the correlations between the recoil $W$ boson and  the associated jets  in 0-30$\%$ central Pb+Pb collisions at the LHC energy within our framework. In our calculations, the only parameter $\alpha_s$ that controls the effective coupling strength between jet and medium is set to 0.2, which is the value we fixed in our previous studies of $Z$+jets correlations~\cite{Zhang:2018urd} in Pb+Pb collisions.

\subsection{Attenuation of $W$-jet in Pb+Pb}

  \begin{figure}

\includegraphics[width=0.56\textwidth]{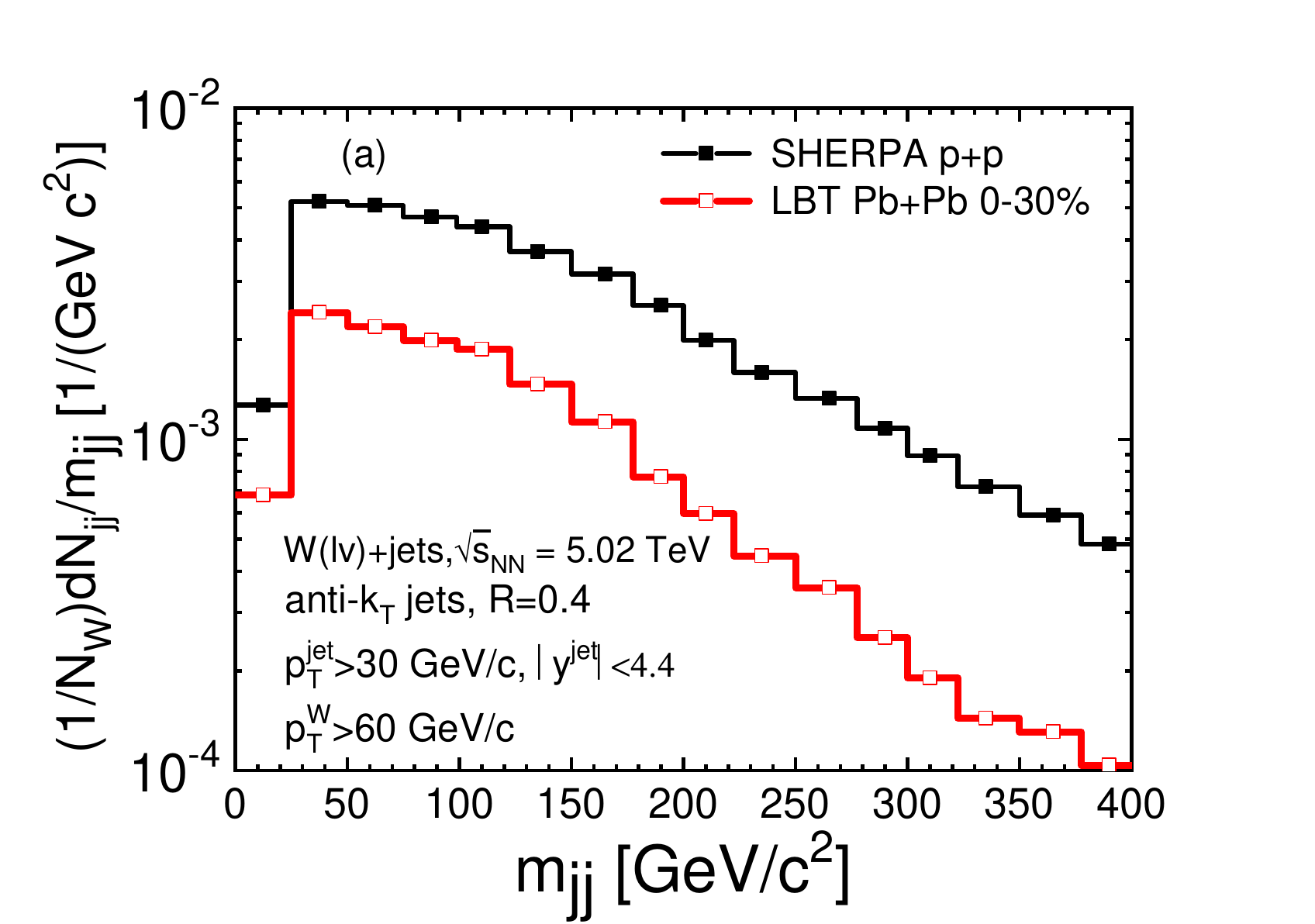}
\includegraphics[width=0.56\textwidth]{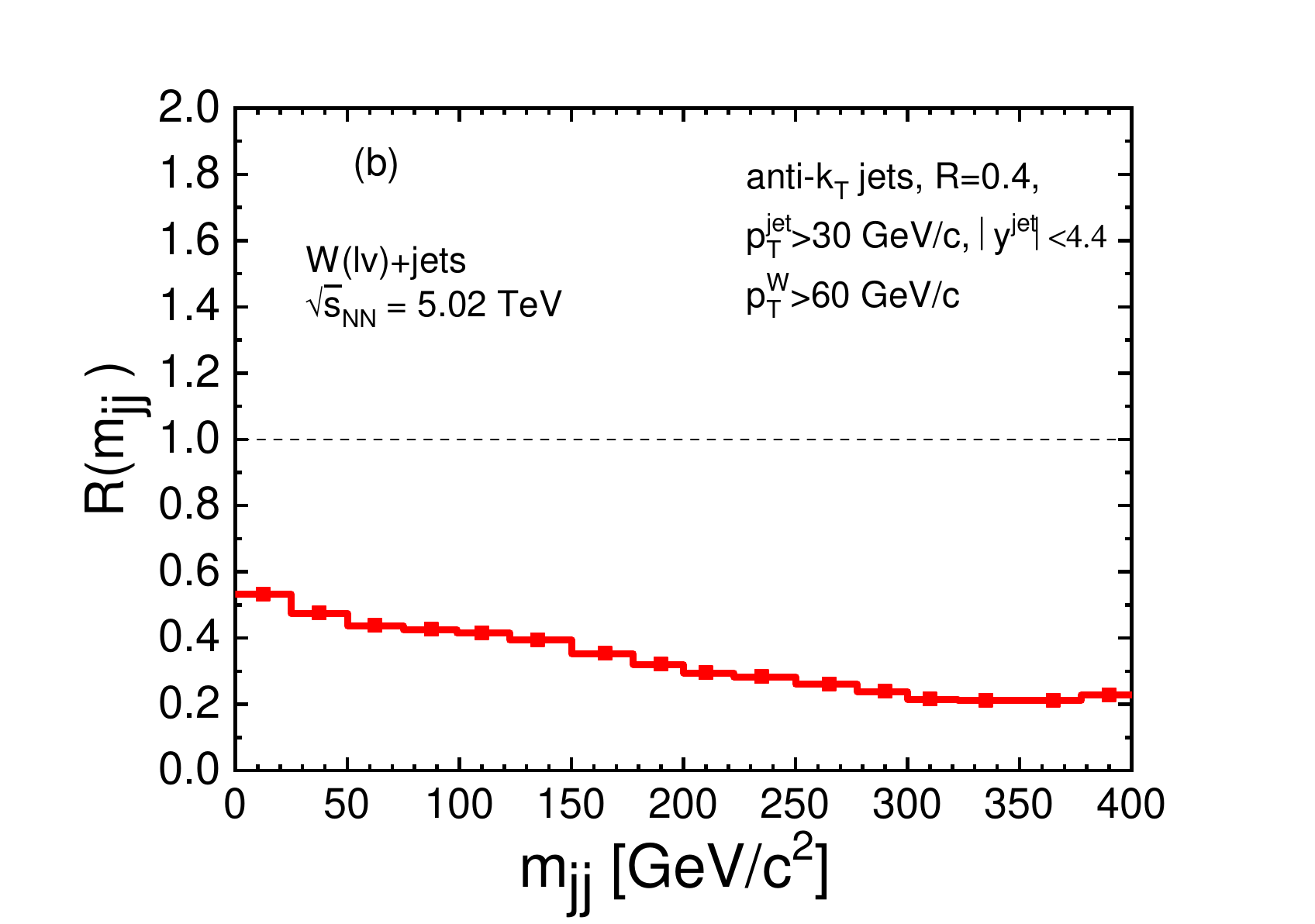}

    \vspace{-0pt}
  \caption{(Color online) (a) Normalized distributions of events passing the $W$ + jets selection cut as a
function of the dijet invariant mass $m_{jj}$ between the two leading jets in $N_{jets}\ge $ 2 events in p+p and  the scaled distributions in Pb+Pb collisions at $\sqrt {s_{NN}}=5.02$ TeV. (b)The ratio of normalized distribution of $m_{jj}$ in Pb+Pb to that in p+p collisions at $\sqrt {s_{NN}}=5.02$ TeV. } \label{mjj}
\end{figure}

We first investigate the distribution in dijet invariant mass $m_{jj}$ between the two leading jets in $W$-jets events with $N_{jets} \ge 2$ and medium modification of the $m_{jj}$ distribution, which is defined as
\begin{equation}
  R_{AA}(m_{jj}) = \frac{1}{\langle N_{coll}\rangle} \frac{dN^{AA}/dm_{jj}}  {dN^{pp}/dm_{jj}}.
\end{equation}

In Fig.~\ref{mjj}, we present the normalized dijet invariant mass distribution for events passing the $W$-jets selection in p+p and the scaled distributions in Pb+Pb collisions at $\sqrt {s_{\rm NN}}=5.02$ TeV as well as the nuclear modification factor. Since the dijet invariant mass is proportional to the virtuality
~\cite{Majumder:2014gda,Connors:2017ptx} of the initial hard scattering, the suppression of the modified invariant mass distribution in Pb+Pb relative to p+p collisions is mainly due to the effects of jet quenching.   We note that the $m_{jj}$ distribution is significantly suppressed due to jet quenching and the modification factor tends to decrease with increasing $m_{jj}$ as shown in Fig.~\ref{mjj}.  The suppression of this dijet invariant mass distribution is due to the reduction of the dijet events  that pass all the selection cuts in Pb+Pb collisions due to jet quenching.  The $m_{jj}$ dependance of the suppression factor also indicates that the effective invariant mass of the dijets that pass the selection cuts is suppressed due to the broadening of each individual jet and their relative momentum.

 \begin{figure}
  \centering
\includegraphics[width=0.56\textwidth]{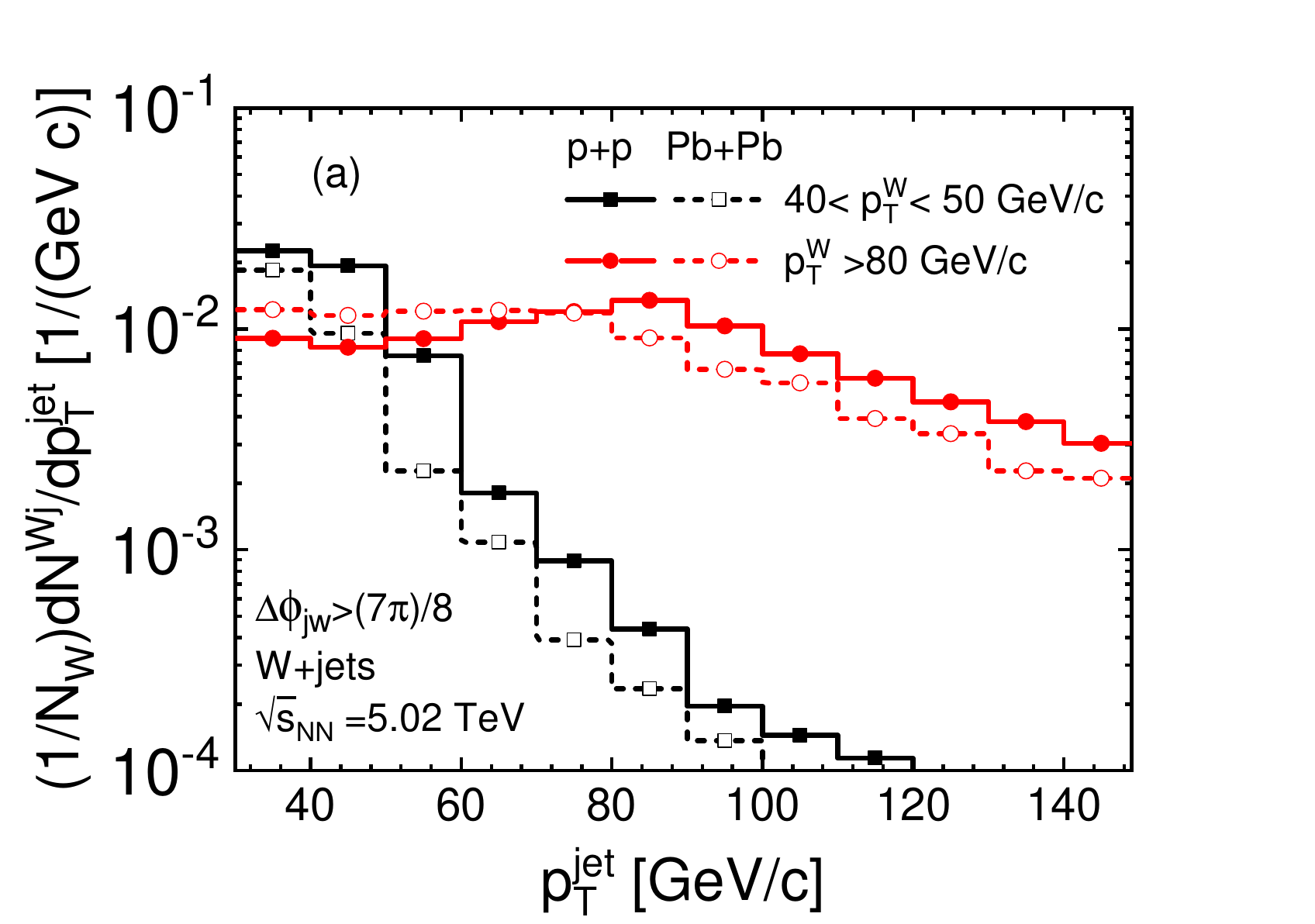}
\includegraphics[width=0.56\textwidth]{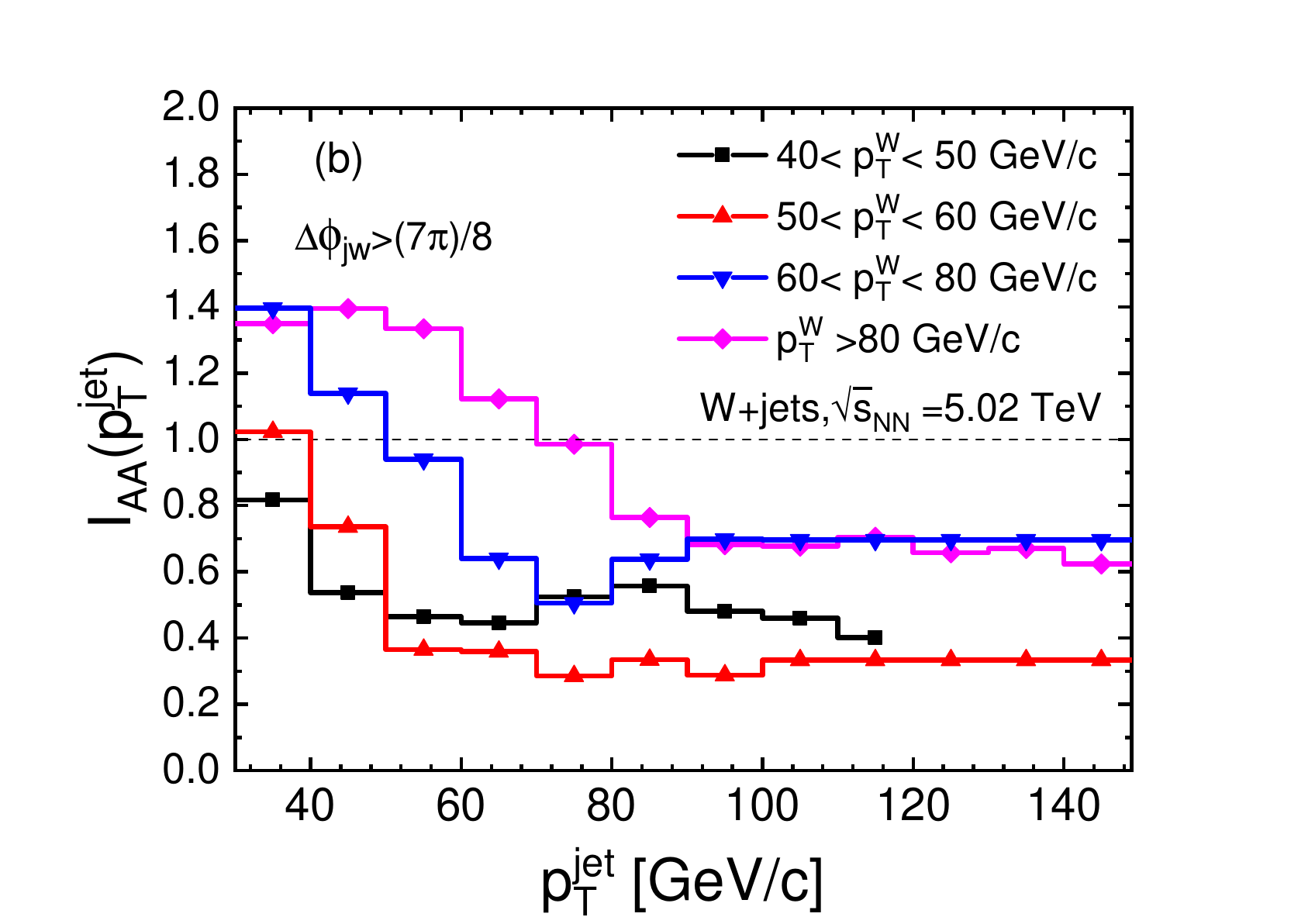}
 \vspace{-10pt}
   \caption{(Color online) (a) The double differential transverse momentum  spectrums of  $W$+jets  in p+p and Pb+Pb collisions in different $p_T^W$ intervals. (b)  Ratio of the transverse momentum of jets associated with a $W$ boson in 0-30$\%$ central Pb+Pb collisions to that in p+p collisions in different transverse momentum ranges of $W$ boson denoted by different color lines, as a function of jet transverse momentum at $\sqrt{ s_{NN} }$=5.02 TeV.}\label{Iaa_W}
\end{figure}

We also calculate another nuclear modification factor for the double differential cross section of W+jet produciton:
   \begin{equation}
   I_{AA}=\frac{(1/N_W^{Pb+Pb})dN^{Pb+Pb}/dp_T^{W}dp_T^{jet}}{ (1/N_W^{p+p})dN^{p+p}/dp_T^{W}dp_T^{jet}},
    \end{equation}
which is defined as the ratio of the double differential tagged jet spectra in central Pb+Pb collisions to that in p+p collisions. The double differential tagged jet spectra in both 0-30\% central Pb+Pb and p+p collisions are shown in  Fig.~\ref{Iaa_W}a and the modification factors are shown in the bottom plot of Fig.~\ref{Iaa_W}b in four $p_T^W$ intervals.

In LO calculations, the jet is produced in the opposite direction of the recoil $W$ boson with the same momentum in the transverse plane. The tagged jet spectra will fall off rapidly above the cutoff value of $p_T^W$. With high-order corrections from  NLO perturbative matrix element calculations of hard emissions  as well as resummation of soft and collinear radiations, the tagged jet spectra are smeared but have a maximum value at around the  $p_T^W$ interval. The jet energy loss in Pb+Pb collisions  will lead to a shift of the tagged jet spectra to a smaller value of $p_T$. This results in the suppression  at low $p_T^{jet}$ and the enhancement  at high $p_T^{jet}$  of the nuclear modification factor $I_{AA}$.  Consequently, the nuclear modification factor is quite sensitive to the transverse momentum cut for the $W$ boson and reach its minimum value in $p_T^{jet}\simeq p_T^W$ region.  This is similar to the jet spectra tagged by direct photon or $Z$ boson.

 \begin{figure}
  \centering
   \includegraphics[width=0.56\textwidth]{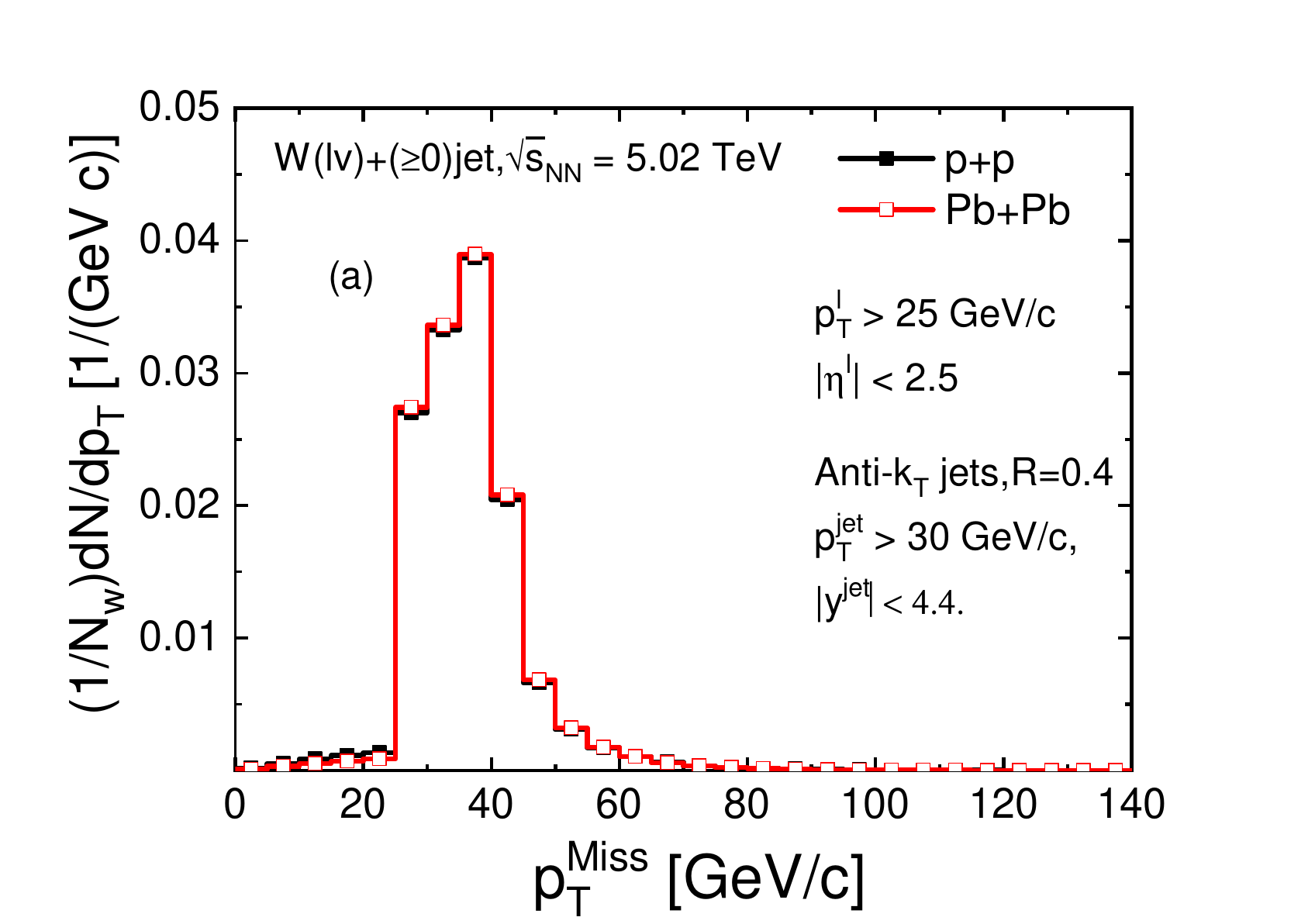}
   \includegraphics[width=0.56\textwidth]{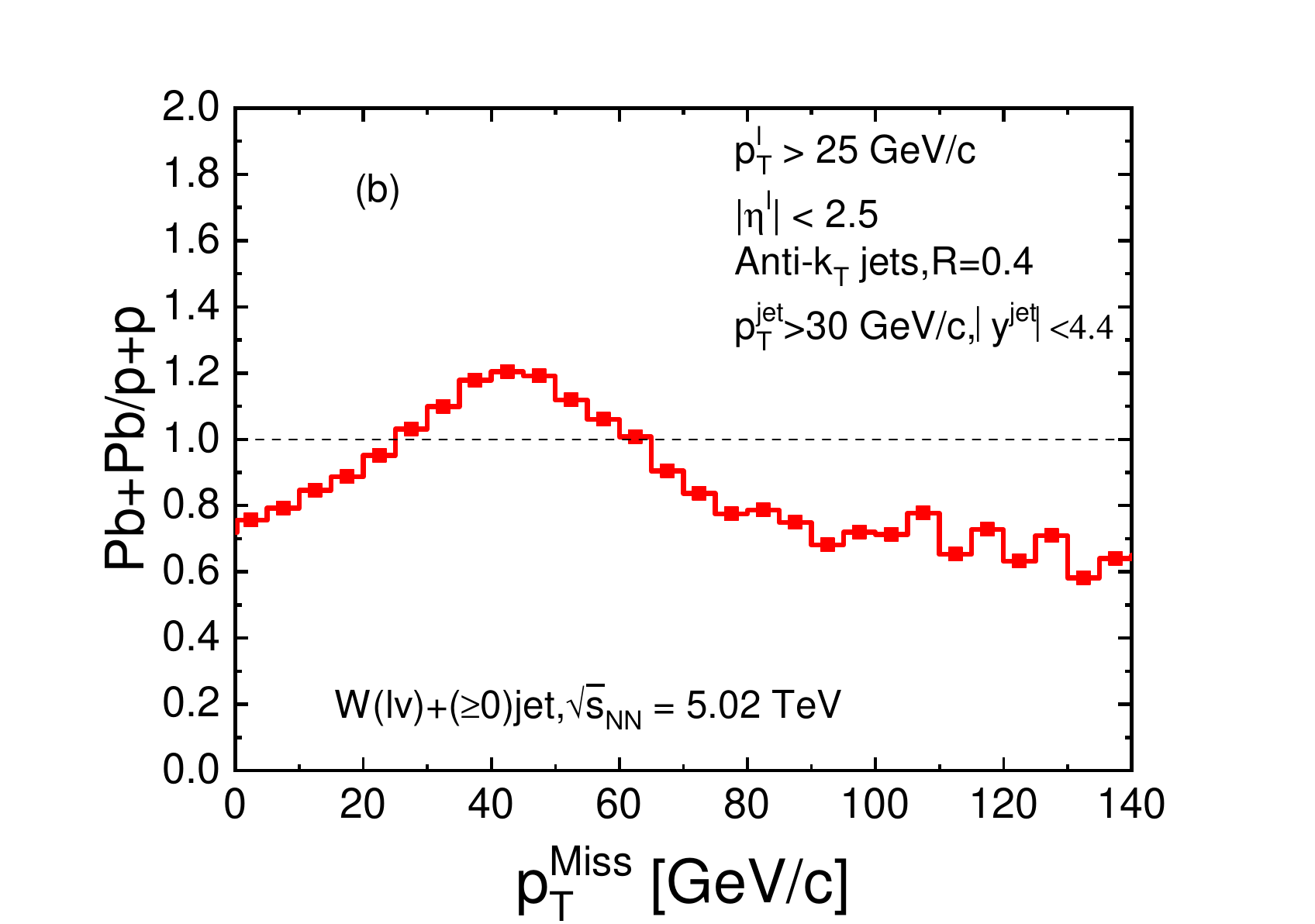}
    \vspace{-0pt}
  \caption{(Color online) (a)  normalized distributions of events passing the $W$ + jets selection cut as a function of the $\vec{p}_T^{Miss}$ which is defined as the vector sum  of the lepton and jets in Pb+Pb and  in p+p collisions at $\sqrt{s_{NN}}=5.02$ TeV. (b):   the ratio of distributions of events passing the $W$ + jets selection as a function of the $\vec{p}_T^{Miss}$ in Pb+Pb to that in p+p collisions at $\sqrt {s_{NN}}=5.02$ TeV.} \label{Mmissing}
\end{figure}

Since the $W$ boson eventually decays into an electron and a neutrino, the existence of the neutrino with missing energy would make the reconstruction of the $W$ boson relatively more difficult than that of $Z^0$ boson, particularly in Pb+Pb collisions with enhanced production of low $p_T$ particles~\cite{Aad:2019sfe}.  When correlation of $W$+jets in heavy-ion collisions is concerned, the situation may be further complicated due to the attenuation of jet energies in the QGP.

To facilitate the experimental study of $W$+jets in Pb+Pb, we define
\begin{equation}
\vec{p}_T^{Miss}=-(\vec{p}_T^{\ l}+ \sum \vec{p}_T^{jets})
\end{equation}
which represents the vector sum of the lepton and jets in a $W$+ jets event, and propose to measure the nuclear modification of
$\vec{p}_T^{Miss}$ distribution as given by:

\begin{equation}
  R_{AA}(p_T^{Miss}) = \frac{1}{\langle N_{coll}\rangle} \frac{ dN^{AA}/dp_T^{Miss}}  {dN^{pp}/dp_T^{Miss}}.
\end{equation}

  \begin{figure}
  \centering
   \includegraphics[width=0.545\textwidth]{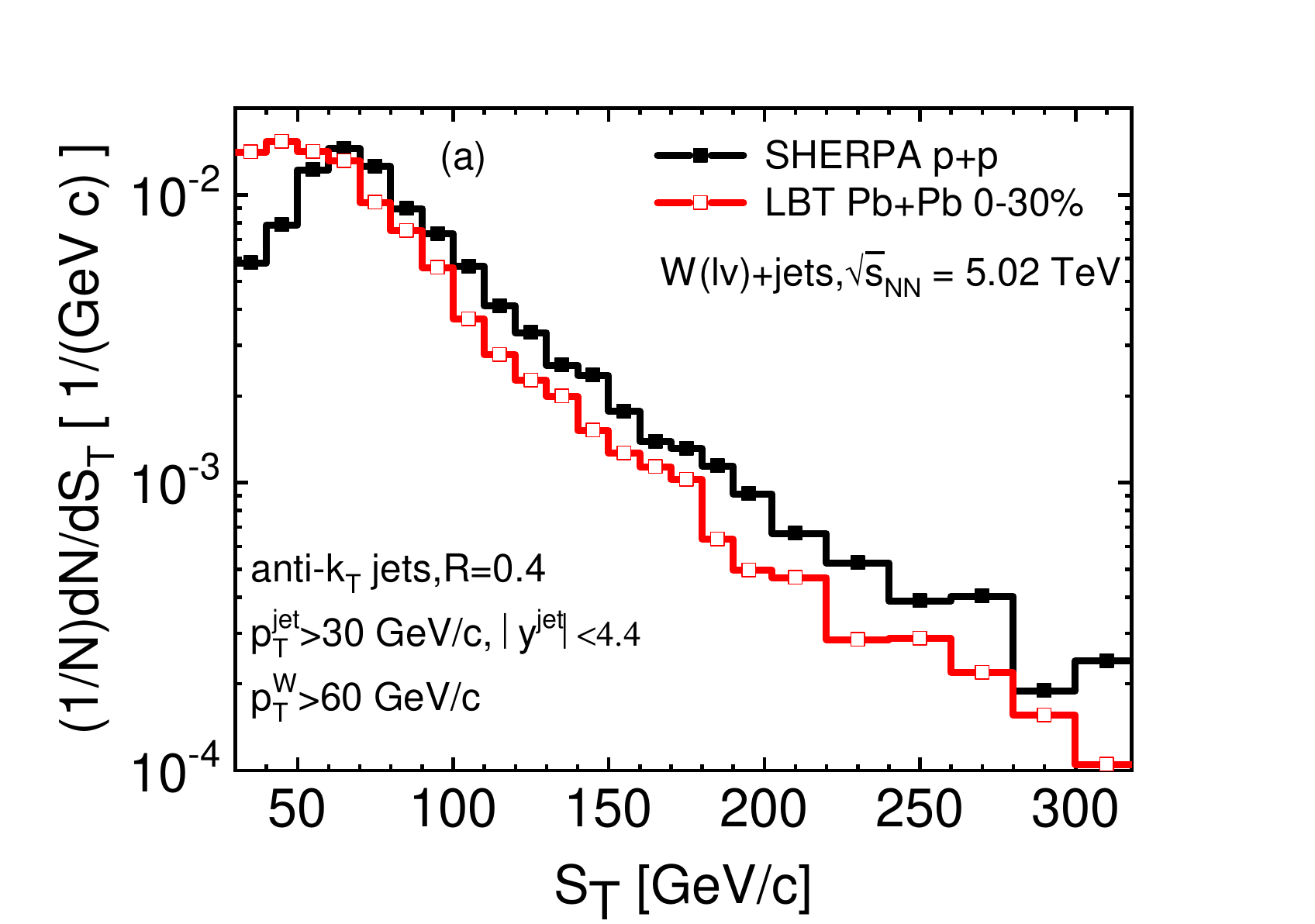}
   \includegraphics[width=0.545\textwidth]{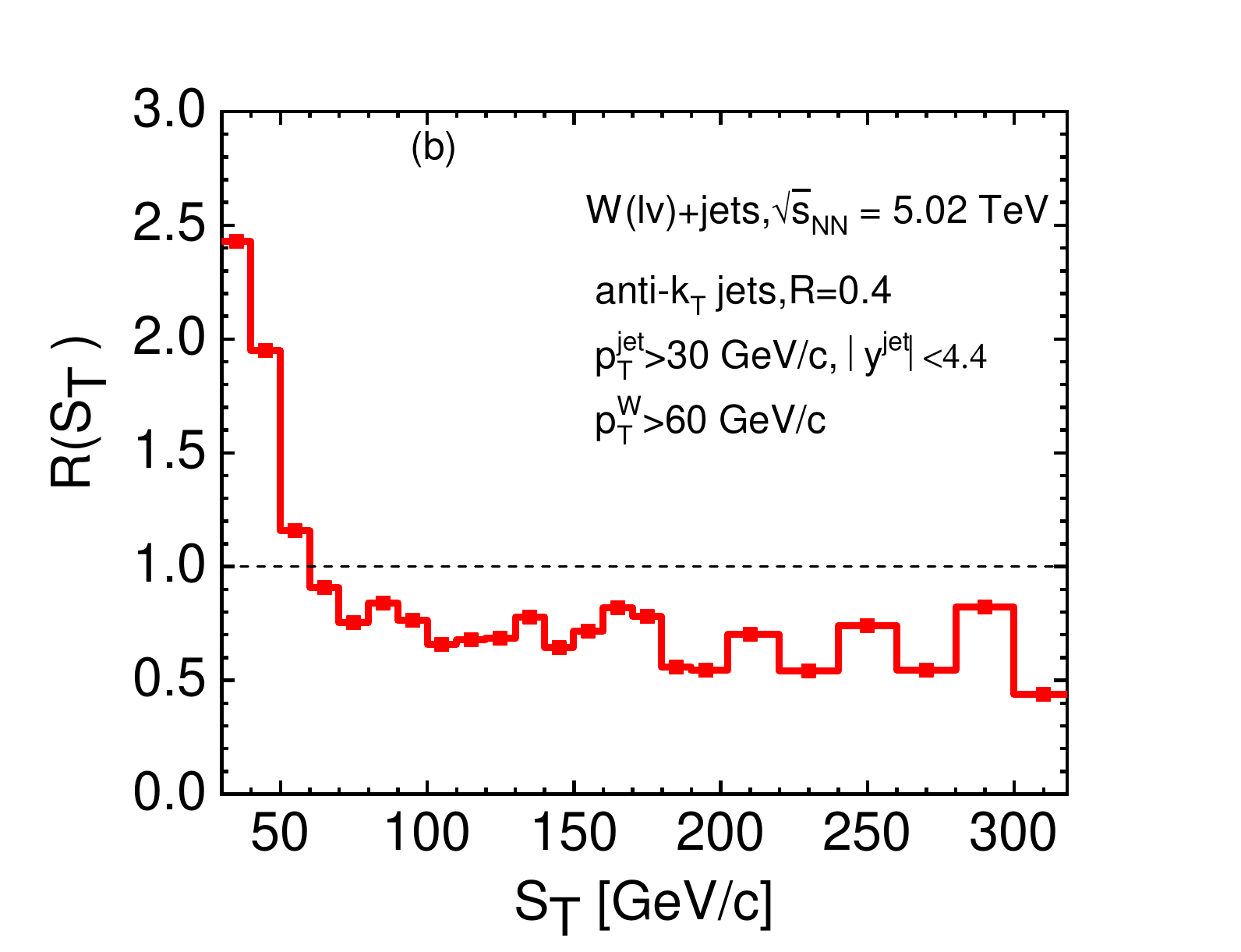}
    \vspace{-10pt}
  \caption{(Color online) (a)the normalized distributions of events passing the $W$ + jets selection as a function of the summed scalar $p_T$ of all reconstructed jets. (b)the ratio of $S_T$ disreibution in Pb+Pb to that in p+p collisions at $\sqrt {s_{NN}}=5.02$ TeV. } \label{SHT}
\end{figure}

The  missing transverse momentum $\vec{p}_T^{Miss}$ excludes neutrino, and only includes lepton and jets.  Therefore it should be much easier to be measured.  In p+p collisions, it is equal to the transverse momentum of the neutrino because of momentum-energy conservation.  In Pb+Pb collisions, it represents the vector sum of the transverse momentum that is outside of the jet cone and the neutrino. This missing energy in Pb+Pb collisions reflects directly the amount and the direction of energy that jets loses in the $W$+jets event in Pb+Pb collisions. The distributions of events passing the $W$ + jets kinematic selection cut as a function of $\vec{p}_T^{Miss}$  in Pb+Pb to that in p+p collisions at $\sqrt s=5.02$ TeV is plotted in  Fig.~\ref{Mmissing}a while their ratio is illustrated in Fig.~\ref{Mmissing}b. One observe that both distributions peak around $p_T^W$, and jet quenching effect in Pb+Pb may shift the peak to a smaller value of $p_T^{miss}$.  This shift is caused by  transverse energy transfer outside the jet cone due to elastic and inelastic scattering with the medium and the direction of the energy carried by radiated partons outside the jet cone  in the opposite direction of the neutrino or $W$ boson. As a consequence, the modification factor $R_{AA}(p_T^{Miss})$ increases as a function of $\vec{p}_T^{Miss}$ in the region $\vec{p}_T^{Miss}<$ 50 GeV/$c$ and decreases with increasing $\vec{p}_T^{Miss}$ in the region $\vec{p}_T^{Miss}>$ 50 GeV/$c$, and is greater than one in the region 30 $<\vec{p}_T^{Miss}<$ 60 GeV/$c$.

To quantify the relative energy loss of $W$+jet events due to jet-medium interaction, we start with the nuclear modification for the summed scalar $p_T$ of all reconstructed jets that pass the kinematic cut in an event $S_T$,  which  should be sensitive to the total transverse momentum broadening of $W$+jets events.

The distributions in Pb+Pb and p+p collisions and the nuclear modification factor,
\begin{equation}
  R_{AA}(S_T) = \frac{1}{\langle N_{coll}\rangle} \frac{dN^{AA}/dS_T}  {dN^{pp}/dS_T},
\end{equation}
as a function of $S_T= \sum p_T^{jets}$ at $\sqrt {s_{NN}}$=5.02 TeV are shown in Fig.~\ref{SHT}.  We note that $R_{AA}(S_T)$ is smaller than one if no cut is adopted on the transverse momentum of the $W$ boson. However, the distributions of $S_T$ is enhanced in the region $S_T<$ 60 GeV/$c$, and suppressed in the region $S_T>$ 60 GeV/$c$ if we adopt a kinematic cut $p_T^W>$ 60 GeV/$c$. $R_{AA}(S_T)$  has similar behaviors as $I_{AA}$ for tagged jet spectra because of the steeply falling cross section when $S_T>p_T^W=$ 60 GeV/$c$. Compared to inclusive jet transverse momentum, the suppression of $R_{AA}$ is a result of the reduction of jet yields as well as the reduction of the jet energy in the QGP. However, $S_T$  is the scalar summed of all the final states jets, the difference of $S_T$  between p+p and Pb+Pb collisions is the total transverse momentum loss or broadening due to jet-medium interactions.

\subsection{Modified correlations of $W$+jet in Pb+Pb}

 \begin{figure}
  \centering
  \includegraphics[scale=0.35]{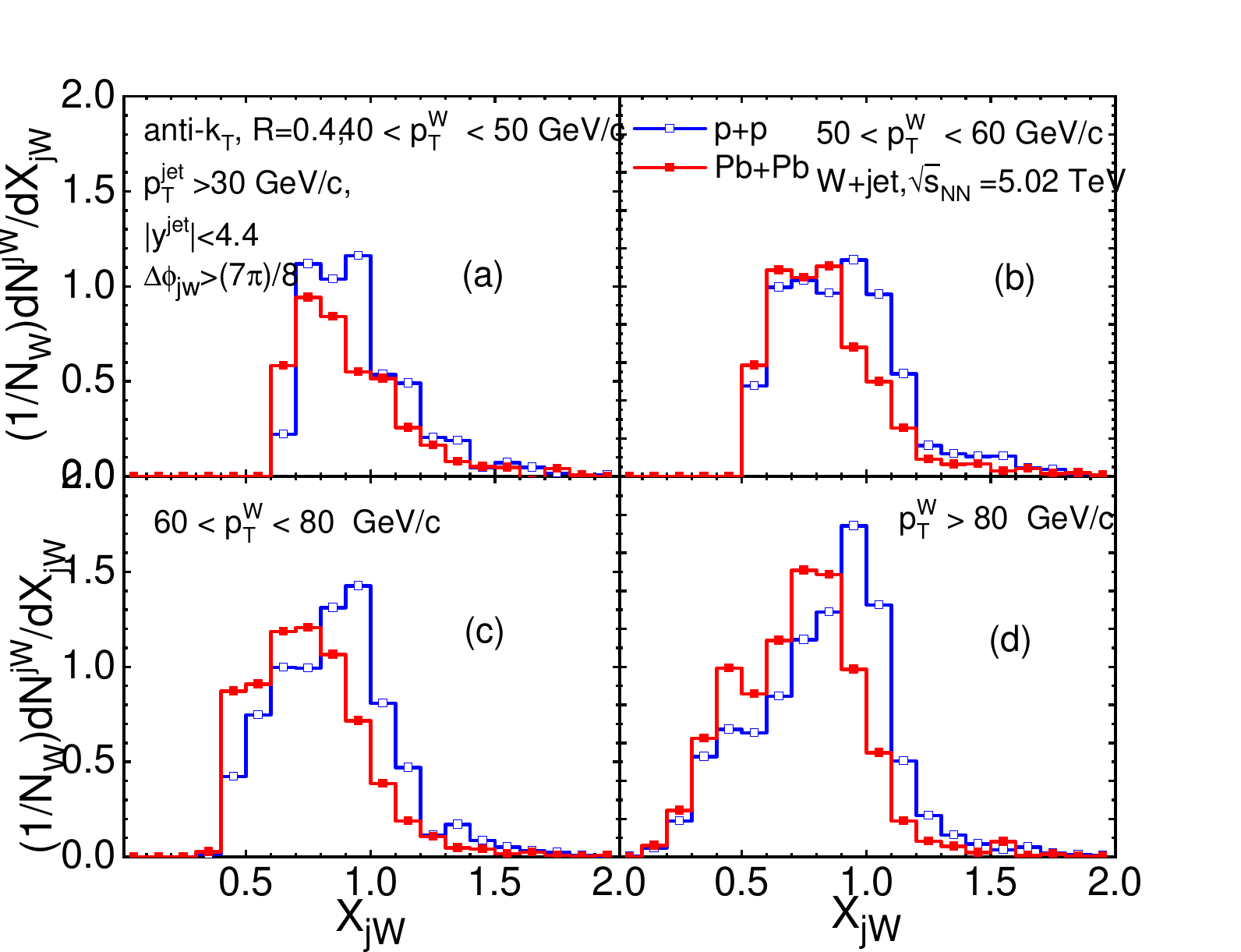}
      \vspace{-5pt}
  \caption{(Color online) Transverse momentum asymmetry $x_{jW}$ of $W$+jets in Pb+Pb and p+p collisions at $\sqrt {s_{NN}}=5.02$ TeV in four $p_T^W$ intervals  (a),(b),(c), and (d). } \label{xjw}
\end{figure}

In this section we turn to correlations between the jet and the recoil $W^{\pm}$ boson. First, the imbalance in the transverse momentum between a $W$ boson and the associated jet $x_{jW}=p_T^{jet}/p_T^W$ is calculated in four $p_T^W$ bins and is shown in Fig.~\ref{xjw}. We only consider the most back-to-back $W$+jet pairs which are required to have azimuthal angle difference $\Delta \phi_{jW}  > 7 \pi /8$.   Even in p+p collisions, the jet energy does not exactly balance the $W$ boson energy because of next-to-leading order effects and some of the quark's energy may extend outside of the jet cone.
Compared to p+p collisions, there is a significant displacement of the peak position of the momentum imbalance $x_{jW}$ towards a smaller value in Pb+Pb collisions.  The shift of the transverse momentum asymmetry is a direct consequence of the energy loss of the jet associated with the $W$ boson with energy above the threshold.  The transverse momentum of the $W$ boson is unattenuated in the QGP,  while the jet loses energy to the outside of the jet cone due to elastic and inelastic interactions with the hot medium constitutions. This leads to a smaller value of $x_{jW}$ in Pb+Pb  compared to that in p+p collisions.

  \begin{figure}
  \centering
   \includegraphics[width=0.56\textwidth]{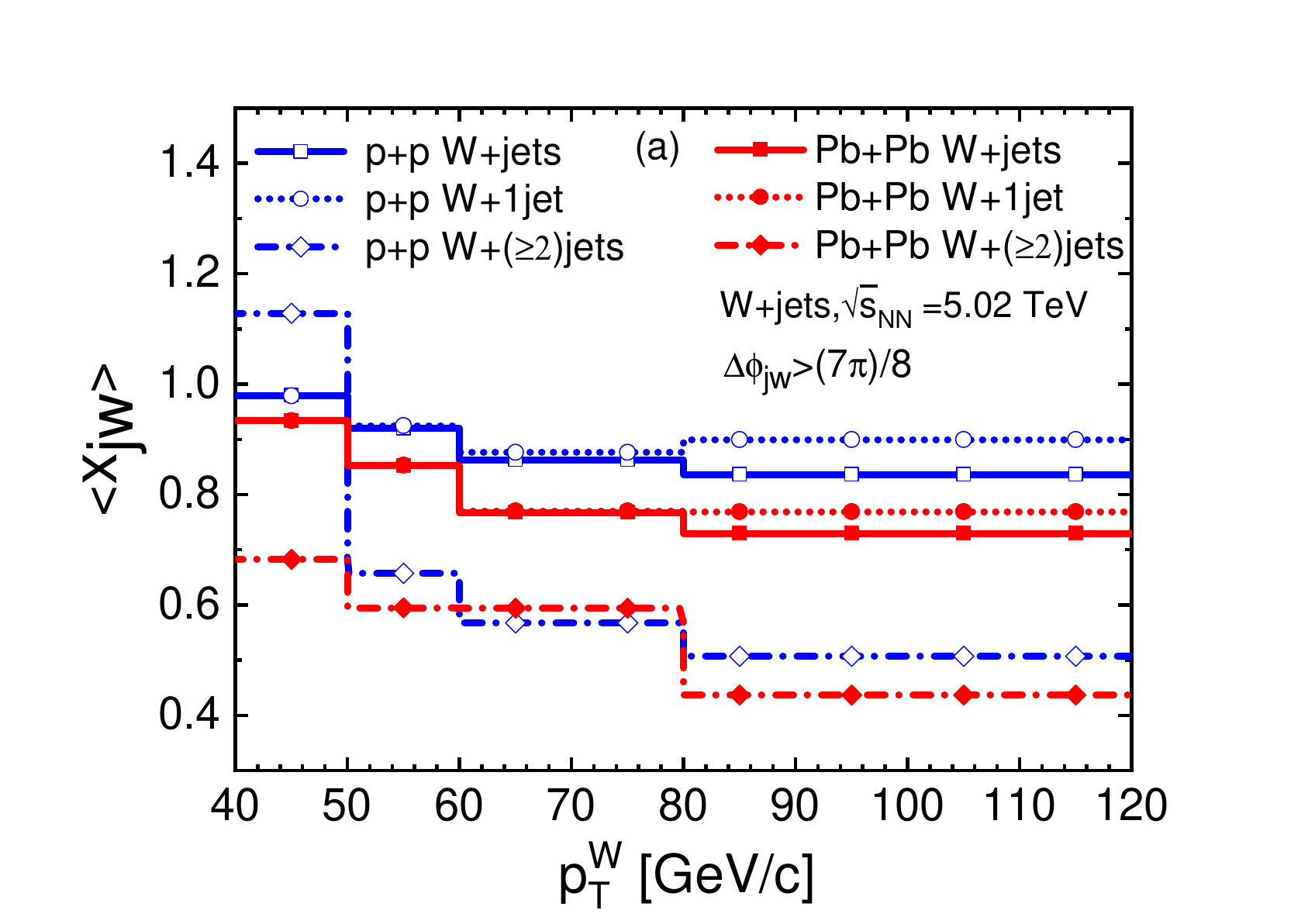}
   \includegraphics[width=0.56\textwidth]{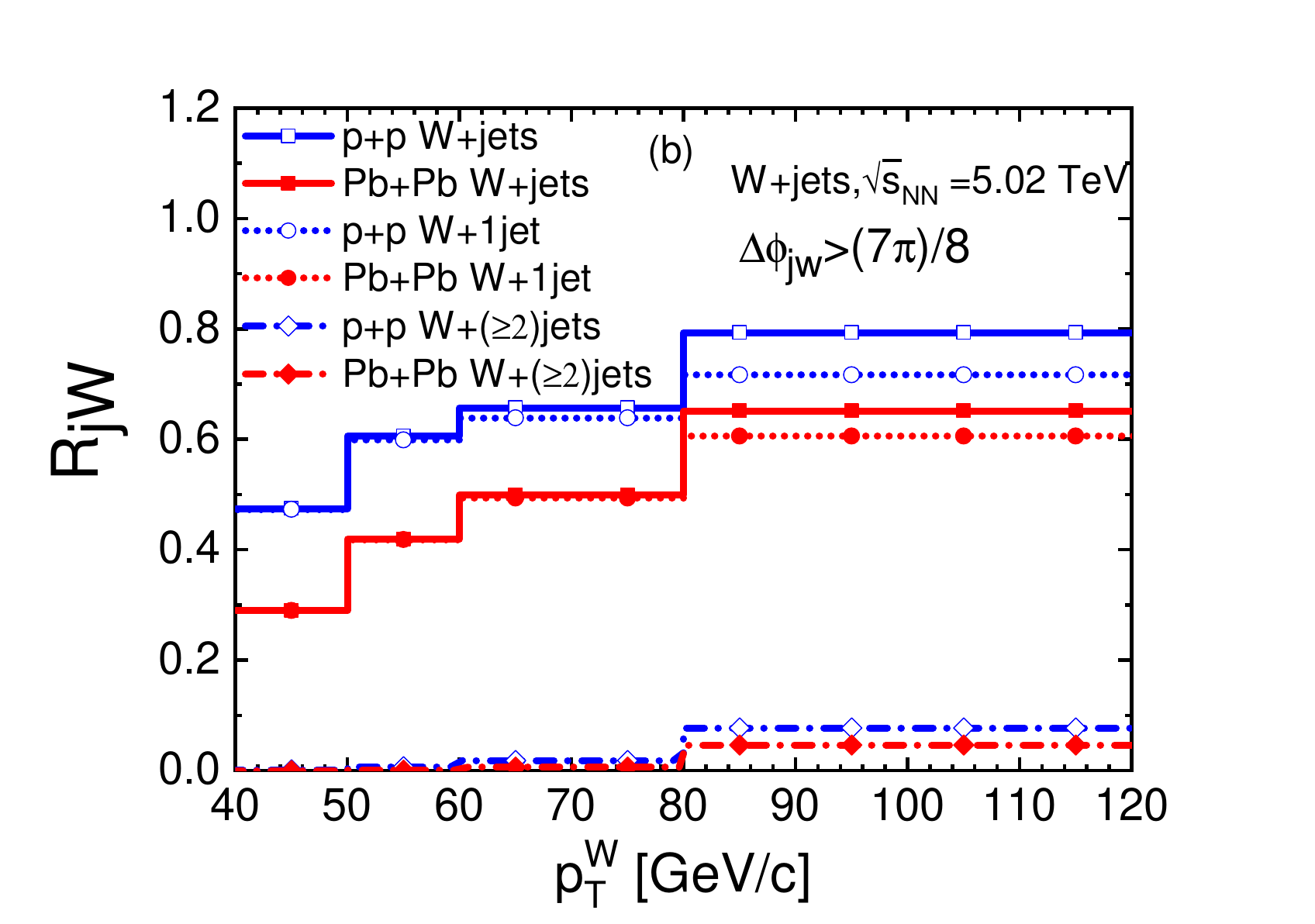}

    \vspace{-0pt}
  \caption{ (Color online) The distribution of (a)the mean value of the momentum imbalance $x_{jW}=p_T^{jet}/p_T^W$ and (b) average number of triggered jets with transverse momentum greater then 30 GeV/$c$ per $W$ boson $R_{jW}$ in Pb+Pb and p+p collisions at $\sqrt {s_{NN}}=5.02$ TeV as a function of the transverse momentum of $W$ boson.}\label{avexjw}
\end{figure}

To quantify the relative shift in the asymmetry distribution between p+p and central Pb+Pb collisions, the mean value of the momentum imbalance $ \langle x_{jW}\rangle$ in different transverse momentum interval of the recoil $W$ boson is calculated as shown in  Fig.~\ref{avexjw}a. We see that the mean value decreases as a function of the transverse momentum of the $W$ boson. When $p_T^W$ is in the interval 40-50 GeV/$c$, the mean value is about 0.98, the energy of the $W$ trigger is almost equal to the  momentum of the associated jet in the transverse plane.  When $p_T^W > 80$ GeV/$c$, the mean value is about 0.85, the jet energy is noticeably smaller than the energy of the recoil  $W$ boson as a result of additional soft or hard emissions from high order corrections. We can also calculate the mean value for a $W$ plus only one jet process (denoted as ``$W$+1jet") and a $W$ in association with more than one jets (denoted as ``$W$+ ($\ge2$)jets")  as shown by the dotted lines and dash dotted lines. We see that the mean value in $W$+($\ge2$)jets processes is about 0.5 for a high energy $W$ boson, indicating that the jet energy in the multi-jet event is only half of the energy of the $W$ boson. The mean value in $W$+($\ge2$)jets processes is greater than 1  when $p_T^W<$ 60 GeV/$c$.  There is a very small probability that a $W$ boson is associated with more than one jets in the back-to-back region with energy greater than $p_T^W$. In those processes, a $W$ boson may be radiated from one of the jets as a fragmentation $W$ boson.

\begin{table}
\begin{center}
\begin{tabular}{|p{2.5cm}<{\centering}|p{1.cm}<{\centering}|p{1.cm}<{\centering}|p{1.cm}<{\centering}|p{1.cm}<{\centering}|}
  \hline

   $p_T^W $(GeV/$c$)& 40-50 & 50-60 & 60-80 &  80-120 \\
  \hline
  $\Delta \langle x_{jW}\rangle$ &0.045&0.068&0.096&0.107 \\
  \hline
    $\Delta \langle x_{jW}\rangle/\langle x_{jW}\rangle_{pp}$ &4.6$\%$&7.4$\%$&11.1$\%$&12.8$\%$ \\
  \hline
  $p_T^W*\Delta \langle x_{jW}\rangle \simeq$ (GeV/$c$)& 2.0&3.7&6.8&10.7 \\
  \hline
\end{tabular}
\caption{ Relative shift of the mean value of momentum imbalance of $W$+jet pair $\langle x_{jW}\rangle$ between p+p collisions and central  Pb+Pb collisions at 5.02 TeV. }\label{xjwdifference}
\end{center}
\end{table}

It is not a surprise that the mean value of the momentum imbalance  in Pb+Pb collisions is much smaller than that in p+p collisions due to jet-medium interactions. The reduction of this mean value in Pb+Pb collisions from that in  p+p collisions $\Delta \langle x_{jW}\rangle$  and fraction of the reduction of the mean value  $\Delta \langle x_{jW}\rangle/\langle x_{jW}\rangle_{pp}$ are tabulated in Table.~\ref{xjwdifference}. We see the reduction increases as a function of the transverse momentum of the $W$ boson. It indicates that jets tagged by higher energy $W$ bosons lose a larger fraction of their energy.

The amount of jet energy loss in Pb+Pb collisions is also shown in the last line in Table~\ref{xjwdifference}. We see that the average jet energy loss increases with the energy of the recoil $W$ boson. With the increased energy of the trigger $W$ boson, the initial transverse momentum of the recoil jet is also larger and it has a higher probability to interact with the medium and loses  larger fraction of its energy to the outside of the jet cone.
However, our results of average jet energy loss in $W+$jets processes is smaller than the Bayesian extraction from $\gamma$+jet~\cite{He:2018gks}. The underlying reason of the difference come from two aspects. First,  the quark fraction in $W+$jets processes is  larger  than that in $\gamma$+jet.  In addition, jet cone size  $R$ used in our calculation is 0.4 while  $R=0.3$ is used in \cite{He:2018gks}.

 Another direct consequence of jet quenching is the reduction of the absolute jet yields above the kinematic threshold in Pb+Pb collisions, which can be investigated through calculating the average number of jet partners per $W$ boson $R_{jW}$.  The dependence of $R_{jW}$ on the transverse momentum of the $W$ boson $p_T^W$ is shown in  Fig.~\ref{avexjw}b. As can be seen, the average number of tagged jets per $W$ boson  that pass the selection cut  is overall suppressed in Pb+Pb due to jet quenching compared to that in p+p collisions.  We also calculated the fraction of jet that fall below the kinematic selection threshold and shown in Table.~\ref{Rjwdifference}. We see that, high energy $W$ bosons lose smaller fraction of jets.
\begin{table}
\begin{center}
\begin{tabular}{|p{2.5cm}<{\centering}|p{1.0cm}<{\centering}|p{1.0cm}<{\centering}|p{1.0cm}<{\centering}|p{1.0cm}<{\centering}|}
  \hline

   $p_T^W $(GeV/$c$)& 40-50 & 50-60 & 60-80 & $\ge$ 80 \\
  \hline
  $\Delta \langle R_{jW}\rangle$ &0.19&0.19&0.16&0.14 \\
  \hline
    $\Delta \langle R_{jW}\rangle/\langle R_{jW}\rangle_{pp}$ &0.39&0.31&0.24&0.18\\
  \hline
\end{tabular}
\caption{Reduction of average number of jet partners per $W$ boson $\langle R_{jW}\rangle$ between p+p collisions and central  Pb+Pb collisions at 5.02 TeV. }\label{Rjwdifference}
\end{center}
\end{table}

  \begin{figure}
  \centering
  \includegraphics[scale=0.35]{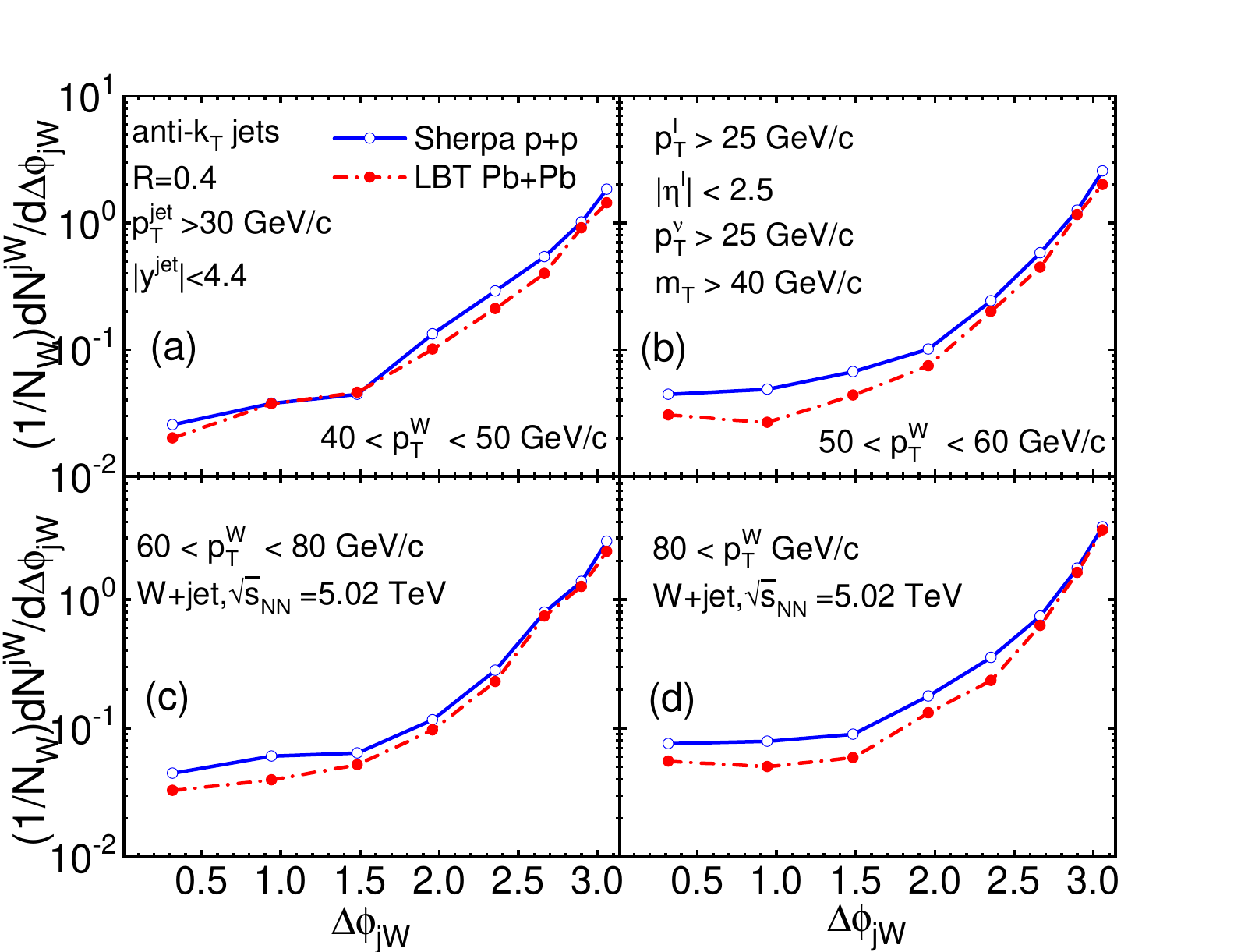}
        \vspace{-5pt}
  \caption{(Color online) Azimuthal angle correlation $\Delta \phi_{jW} $ of $W$+jets in Pb+Pb and p+p collisions at $\sqrt {s_{NN}}=5.02$ TeV in four $p_T^W$ intervals (a),(b),(c), and (d).  } \label{phijw}
\end{figure}

In addition to the transverse momentum correlations, we also calculate the azimuthal angle correlation of jets and the recoil $W$ boson in four $p_T^W$ intervals in both p+p and Pb+Pb collisions as shown in Fig.~\ref{phijw}.  Compared to p+p collisions, the correlation is moderately suppressed at small azimuthal angle (relative to the $W$ boson) in Pb+Pb collisions. The distribution is normalized to the number of $W$ bosons that pass the selection cut rather than the number of $W$+jet pair, and one boson may can not find any associated jets or can have more than one associated jets,  so the integration of the azimuthal angle correlation may be less or greater than one. The suppression of the small angle correlation of $W$+jets is mainly due to the suppression of the secondary or  multiple jets by jet quenching similar to the $\gamma$/$Z$+jets correlation \cite{Luo:2018pto,Zhang:2018urd}.

\begin{figure}
  \centering
   \includegraphics[width=0.56\textwidth]{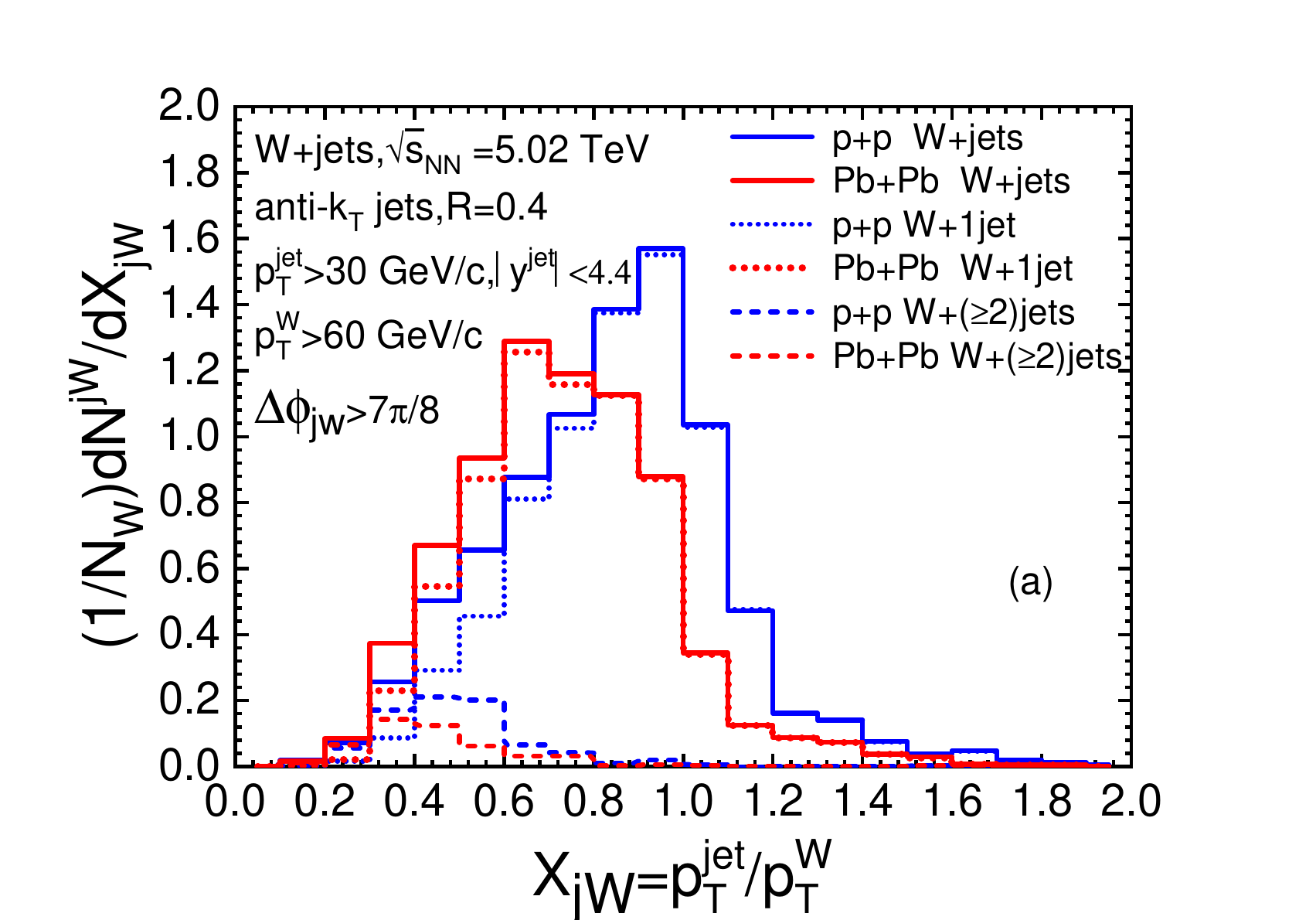}
   \includegraphics[width=0.56\textwidth]{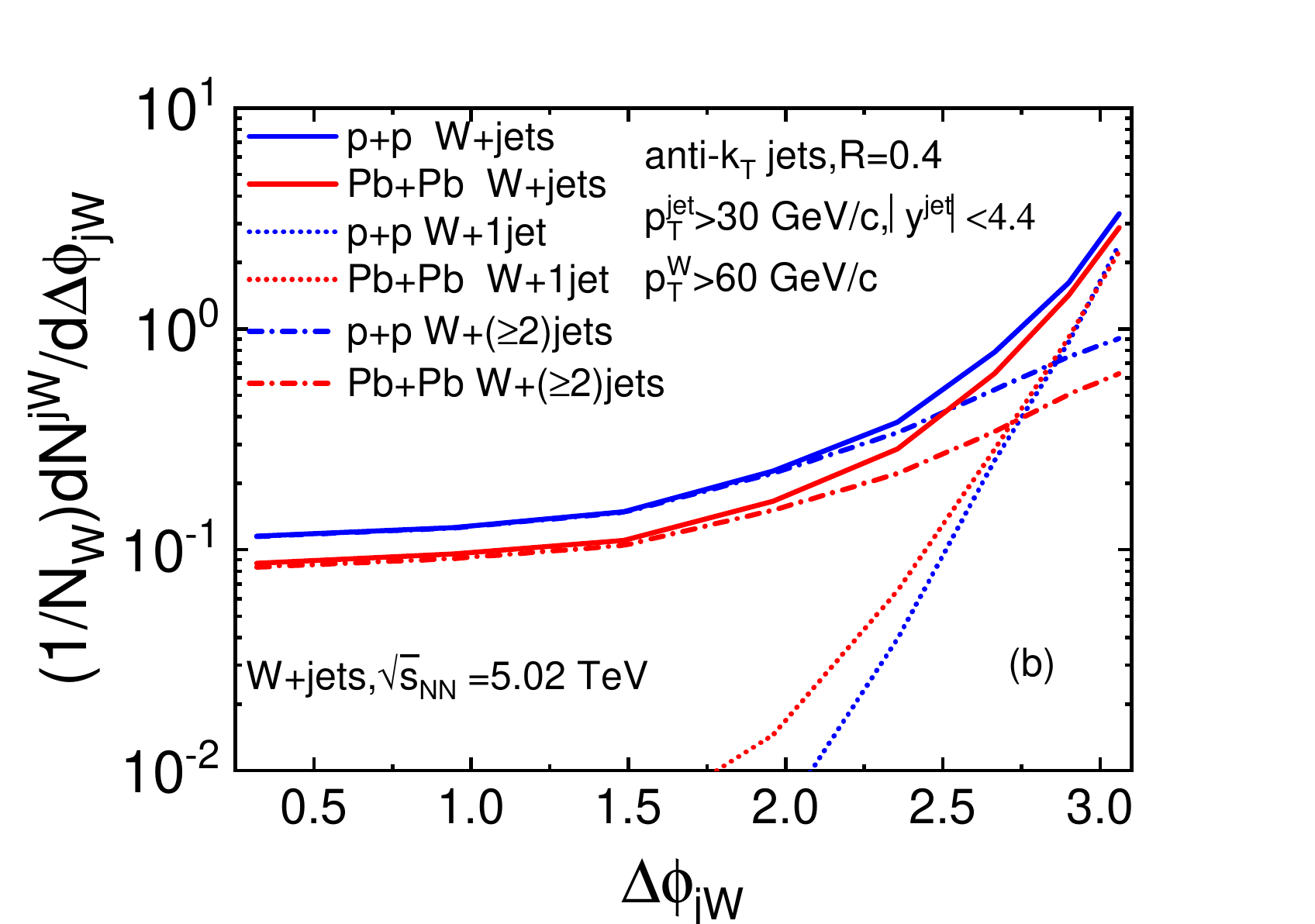}
    \vspace{-0pt}
  \caption{(Color online) (a) Transverse momentum asymmetry $x_{jW}$ and (b) azimuthal angle correlation $\Delta \phi_{jW} $  of $W$+jets in Pb+Pb and p+p collisions at $\sqrt {s_{NN}}=5.02$ TeV.  The contributions from $W$ plus only one jets and $W$ associated with more one jets to $\Delta \phi_{jW} $ and $x_{jW}$ are calculated and shown by dotted line and dash-dotted line respectively.} \label{phijw60}
\end{figure}

To illustrate the mechanism of this suppression, the contributions from $W$ plus only one jet and $W$ in association with  more than one jet to the  transverse momentum asymmetry  and azimuthal angle correlation are also calculated as shown by dotted line and dash-dotted line both in p+p and Pb+Pb collisions respectively in Fig.~\ref{phijw60}.  We see that $W$+1jet processes dominate the large angle region where the jet  is opposite to the direction of the $W$ boson in the transverse plane. However, in the small angle region, it is the $W$ plus more than one jets processes that dominate the correlation. Compared to $W$ plus only one jet, the azimuthal angle correlation of $W$ associated with more than one jets is much broadened. This is because, $W$ plus only one jet processes mainly come from leading-order matrix element and the $W$ boson is balanced by only one jet with azimuthal angle around $180^0$ relative to the recoil $W$ boson.   On the other hand, $W$ production associated with more than one jet mainly originates from  NLO matrix elements which contain hard emissions at large angles. These multiple jets with relatively lower energy can easily lose energy due to jet quenching and shift their final energy below the kinematic cut.  This leads to the suppression of $W$+jets correlations at small azimuthal angle.

  \begin{figure}
  \centering
   \includegraphics[width=0.56\textwidth]{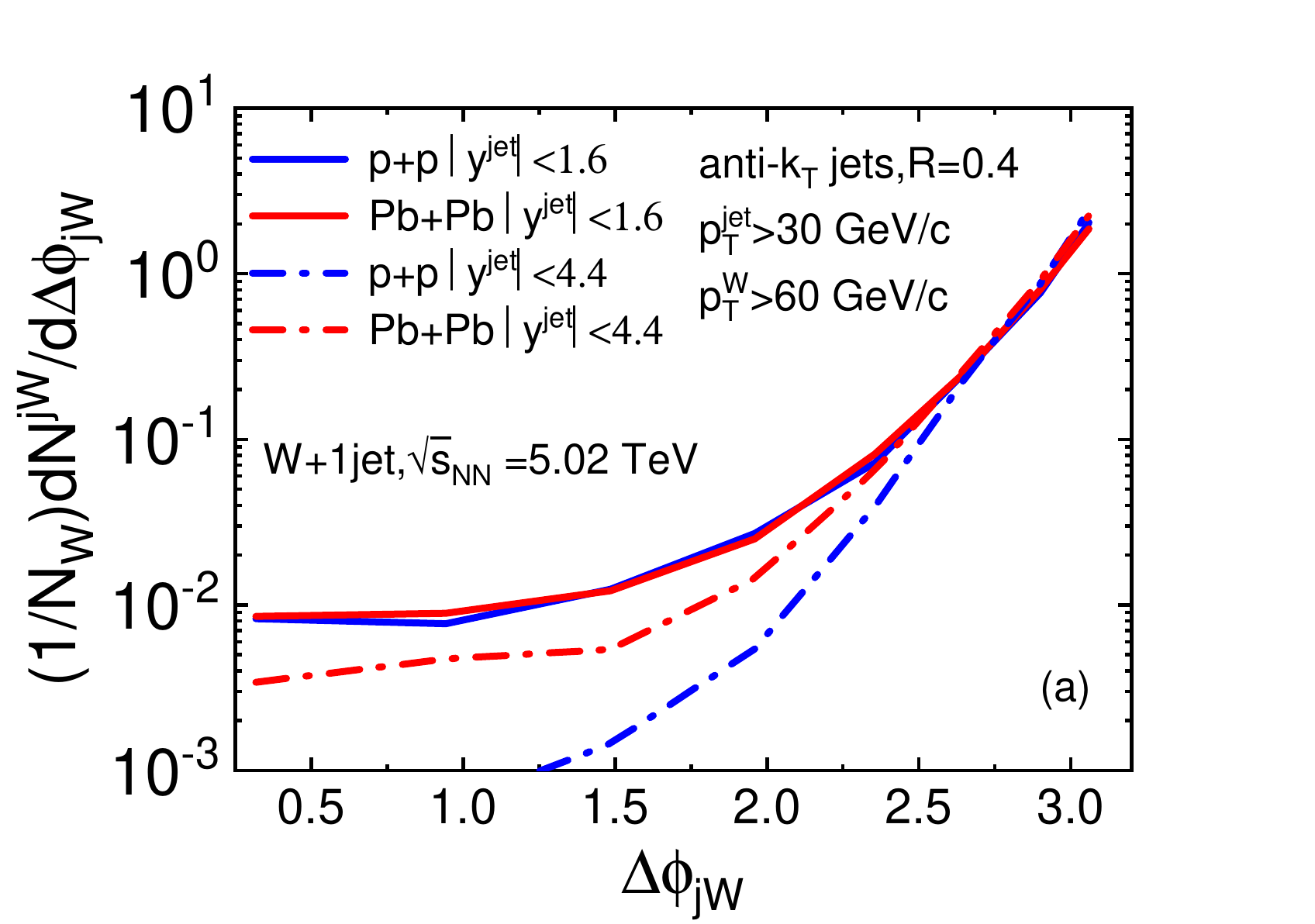}
   \includegraphics[width=0.56\textwidth]{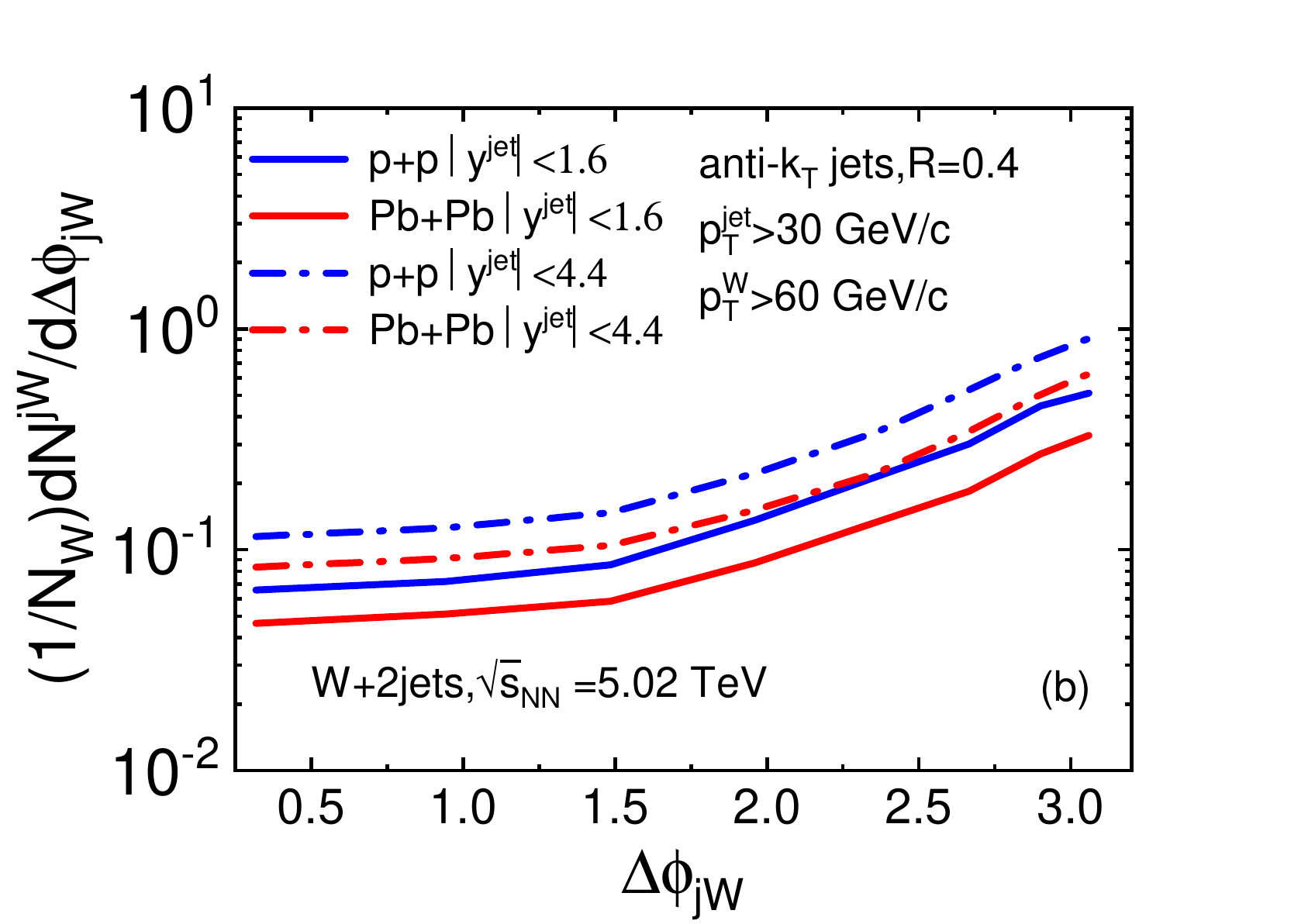}
    \vspace{-0pt}

  \caption{(Color online) Azimuthal angle correlation $\Delta \phi_{jW} $ (a) $W$ plus only one jets and (b) $W$ associated with more one jets in Pb+Pb and p+p collisions at $\sqrt {s_{NN}}=5.02$ TeV. The  $\Delta \phi_{jW} $  distributions with $|y|$ cut 1.6 and 4.4 are shown by solid line and dash-dotted line respectively.} \label{phiwj}
\end{figure}

We observe a moderately  broadened  W-jet correlation in these events in Pb+Pb relative to p+p collisions, which is different with the results of $Z$+jets~\cite{Zhang:2018urd} where no modification is observed in $Z$ plus only one jet process between p+p and Pb+Pb collisions. The underlying reason of the difference is that the kinematic threshold used in those calculations is different, especially the cut on the jet rapidity.  Fig.~\ref{phiwj} shows the azimuthal angle correlation $\Delta \phi_{jW} $ of  $W$ plus only one jets (up) and $W$ associated with more one jets (bottom) with jet rapidity $|y^{jet}|<1.6$ and $|y^{jet}|<4.4$ and the comparison between p+p and Pb+Pb collisions. We see  $W$+1jets is much broader with jet rapidity $|y^{jet}|<1.6$ compared to $|y^{jet}|<4.4$,  while contributions from $W$+2jets is much enhanced with jet rapidity $|y^{jet}|< 4.4$ compared to $|y^{jet}|<1.6$.   This is because larger rapidity cut would include more jets, as a result of which, the  $W$ event in association with only one jet with constraint $|y^{jet}|<1.6$ would become a event in which  $W$ is  associated with more than one jet with condition $|y^{jet}|<4.4$. With constraint $|y^{jet}|<1.6$, no significant difference of the $\Delta \phi_{jW} $ in  $W$+1jet is observed between p+p and Pb+Pb collisions as in  $Z$+jets~\cite{Zhang:2018urd}.  However,  a moderately  broadened  W-jet correlation is seen with constraint $|y^{jet}|<4.4$ in Pb+Pb compared to p+p collisions.  This is because the energy of the jets with large rapidity in $W$ plus multi-jets events is relative small. When some of those jets  lose energy and get easily lost in Pb+Pb collisions.  Some of these $W$ plus multi-jets events in p+p collisions would become $W$+1jet process in Pb+Pb collisions due to jet quenching. This leads to the enhancement of $W$+1jet azimuthal correlation in the small angle region with kinematic constraint $|y^{jet}|<4.4$ in Pb+Pb relative to p+p collisions.

\section{Conclusion}
\label{sec:conclusion}

We have carried out the first systematic study of jet production in association with a $W$ boson in both p+p and Pb+Pb collisions at the LHC energy.  We use a Monte Carlo event generator SHERPA to generate reference $W$+jet  production in p+p collisions with NLO ME matched to PS. Our calculations show excellent agreement with the experimental data in p+p collisions.  Jet propagation and medium response in the hot/dense medium are simulated by LBT and the medium information is provided by 3+1D CLVisc hydrodynamics. We  investigated the medium effect on the dijet invariant mass $m_{jj}$ between the two leading jets. We also studied the nuclear modification of jet spectra associated with a $W$ boson in different $W$ transverse momentum intervals. Jet-medium interactions lead to an enhancement in small $p_T^{jet}$ region and  a suppression in large $p_T^{jet}$ region due to the steep falling of the jet spectra.  We also presented the modification of the  missing transverse momentum in $W$+ jets events. The shift of this distribution to a smaller value indicates that jets lose large fraction of their energy in the opposite direction of the neutrino or $W$ boson. We demonstrate that the shift of the scalar sum of transverse momentum $S_T$ reflects the absolute jet energy loss in Pb+Pb collisions. Furthermore, we have investigated the shift of  $W$+jet $p_T$ imbalance distribution $x_{jW}$ due to jet energy loss, the suppression of jet partners per $W$ trigger $R_{jW}$ due to the reduction of jets yields, as well as the modification of $W$+jet azimuthal angle correlations $\Delta \phi_{jW}$ resulting from the suppression of multi-jets in heavy-ion collisions.


{\bf Acknowledgments:}  The authors would like to thank H Zhang, T Luo, P Ru, G Ma for helpful discussions. This research is supported by Guangdong Major Project of Basic and Applied Basic Research No. 2020B0301030008, Natural Science Foundation of China with Project No. 11935007, 11805167.

\end{document}